\documentclass{aa}
\usepackage{graphicx,txfonts}

\begin{document}

\title{
	Oxygen trends in the Galactic thin and thick disks\thanks{Based on 
	observations collected at the European Southern Observatory, La Silla 
	and Paranal, Chile, Proposals \#65.L-0019, 67.B-0108, and 69.B-0277}
	\fnmsep\thanks{The full Table~\ref{tab:o7774abundances} is only 
	available in electronic form at the CDS via anonymous ftp to 
	\texttt{cdsarc.u-strasbg.fr} (130.79.128.5) or via
        \texttt{http://cdsweb.u-strasbg.fr/cats/J.A+A.all.htx}}
	}

\titlerunning{
	Oxygen trends in the Galactic thin and thick disks
	}

\author{
	T. Bensby
	\and S. Feltzing
	\and I. Lundstr\"om
	}

\offprints{Thomas Bensby, \texttt{thomas@astro.lu.se}}

\institute{Lund Observatory, Box 43, SE-221\,00 Lund, Sweden}

\date{Received 21 July 2003/ Accepted 24 October 2003}

%==============================================================================
\abstract{
We present oxygen abundances for 72 F and G dwarf stars in the solar 
neighbourhood. Using the kinematics of the stars we divide them into two 
sub-samples with space velocities that are typical for the thick and thin 
disks, respectively. The metallicities of the stars range from
[Fe/H]\,$\approx$\,$-0.9$ to $+0.4$ and we use the derived oxygen abundances 
of the stars to: 
(1) perform a differential study of the oxygen trends in the thin 
and the thick disk; 
(2) to follow the trend of oxygen in the thin disk to the highest 
metallicities. 
We analyze the forbidden oxygen lines at 6300\,{\AA} 
and 6363\,{\AA} as well as the (NLTE afflicted) triplet lines around 
7774\,{\AA}. For the forbidden line at 6300\,{\AA} we have spectra of very high 
$S/N$ ($>$400) and resolution ($R$\,$\gtrsim$\,215\,000). This has enabled a 
very accurate modeling of the oxygen line and the blending Ni lines. The high 
internal accuracy in our determination of the oxygen abundances from this line 
is reflected in the very tight trends we find for oxygen relative to iron. 
From these abundances we are able to draw the following major conclusions: 
(i) That the [O/Fe] trend at super-solar [Fe/H] continues downward which is in
concordance with models of Galactic chemical evolution. This is not seen 
in previous studies as it has not been possible to take the blending Ni lines 
in the forbidden oxygen line at 6300\,{\AA} properly into account;
(ii) That the oxygen trends in the thin and the thick disks are distinctly
different. This confirms and extends previous studies of the other 
$\alpha$-elements;
(iii) That oxygen does not follow Mg at super-solar metallicities;
(iv) We also provide an empirical NLTE correction for the infrared \ion{O}{i} 
triplet that could be used for dwarf star spectra with a $S/N$ such that only
the triplet lines can be analyzed well, e.g. stars at large distances;
(v) Finally, we find that Gratton et al. (1999) overestimate the 
NLTE corrections for the permitted oxygen triplet lines at $\sim$\,7774\,{\AA} 
for the parameter space that our stars span.
	}

\maketitle

%==============================================================================
\section{Introduction}

Oxygen is, next to hydrogen and helium, the most abundant element in the 
Universe. Of the three stable isotopes $^{16}$O, $^{17}$O, and $^{18}$O, the 
first is the dominating one, making up $\sim$\,99.8\,\% of the total oxygen 
content in the Solar system. Oxygen is a bona fide primary element that 
essentially only forms in the interiors of massive stars through hydrostatic 
burning of mainly He, C, and Ne. By analyzing elemental abundances in the 
atmospheres of long-lived F and G dwarf stars it is not only possible to 
determine the chemical composition of the gas that the stars were born out of 
but also to trace the chemical history in the different stellar populations in 
the Milky Way. In this respect the oxygen fossil record is of extra importance 
and is often used in models of Galactic evolution. It can, among other things, 
be used to measure the rates of supernovae type~II (SN\,II) and supernovae 
type~Ia (SN\,Ia) with time (e.g. Wheeler at al.~\cite{wheeler}). An 
overabundance of oxygen ([O/Fe]\footnote{Abundances expressed within brackets 
are as usual relative to solar values:
$[{\rm O}/{\rm Fe}] = \log (N_{\rm O} / N_{\rm Fe})_{\star} - 
                      \log (N_{\rm O} / N_{\rm Fe})_{\sun}$}\,$>$\,0)
indicates that the region have had a high star formation rate and undergone a 
fast chemical enrichment (e.g. Tinsley~\cite{tinsley}; 
Matteucci \& Greggio~\cite{matteucci}).

The determination of oxygen abundances is unfortunately often troublesome due 
to the limited number of available oxygen lines in the visual part of a stellar 
spectrum. The main indicators are the \ion{O}{i} triplet at 
$\sim$\,7774\,{\AA}, the forbidden [\ion{O}{i}]\footnote{These brackets 
indicate that the spectral line is forbidden and should not be confused with 
the notation of abundances given relative to solar values.} lines, especially 
those at 6300\,{\AA} and 6363\,{\AA}, the ultraviolet (UV), and the infrared 
(IR) OH lines. The analysis of these lines all have their difficulties. The 
triplet at 7774\,{\AA} should be ideal to work with since its lines are strong 
and are located in a clean part of the spectrum that is free from blending 
lines. The lines are, however, strongly affected by deviations from local 
thermal equilibrium (LTE) (see e.g. Kiselman~\cite{kiselman}) and, even if 
these deviations are considered not very well understood even for the Sun,
they are probably due to convective inhomogeneities (Kiselman~\cite{kiselman2}). 
The forbidden lines are very robust indicators but are hard to work with since 
they are both weak (the measured equivalent widths in the Sun are 
$W_{\lambda}(6300)$\,$\approx$\,5\,m{\AA} and 
$W_{\lambda}(6363)$\,$\approx$\,3\,m{\AA}, Moore et al.~\cite{moore}).
Both lines are also blended by lines from other species. 
The [\ion{O}{i}] line at 6300\,{\AA} has two \ion{Ni}{i} lines in its right 
wing (Lambert~\cite{lambert}; Johansson et al.~\cite{johansson}) and the 
[\ion{O}{i}] line at 6363\,{\AA} has a possible CN contribution 
(Lambert~\cite{lambert}).

The conclusion from the Joint Discussion 8 during the IAU General Assembly in 
Manchester in 2000 (New Astronomy Reviews 45, 2001), which was devoted to a 
discussion of oxygen abundances in old stars, is that the most reliable 
indicators of stellar oxygen abundances are the forbidden [\ion{O}{i}] lines 
(Barbuy et al.~\cite{barbuy2}). The [\ion{O}{i}] line at 6300\,{\AA}
is blended and is located in a part of the spectrum that is severely affected
by telluric lines. If we want to use this line for abundance analysis
it is important to be able to take these effects
into account properly. This is best done by using spectra with high 
signal-to-noise ratios ($S/N$) and high resolutions ($R$) together with 
accurate atomic line data, $\log gf$-values in particular.

In the Milky Way it is mainly the halo stellar population that shows the large 
overabundances of oxygen relative to iron, indicative of a fast star formation. 
There is however an ongoing controversy about the size of this overabundance. 
Depending on which indicators that are used in the determination of the oxygen 
abundances one can find different [O/Fe] trends for these stars. The forbidden 
oxygen line at 6300\,{\AA} and the IR OH lines generally produce a constant 
value of [O/Fe]\,$\approx$\,0.5 for halo stars with [Fe/H]\,$\lesssim$\,$-1$, 
see e.g. Nissen et al.~(\cite{nissen2002}), 
Gratton \& Ortolani~(\cite{gratton1986}), 
Sneden et al.~(\cite{sneden}), and
Lambert et al.~(\cite{lambert1974}) for abundances from the [\ion{O}{i}] line 
at 6300\,{\AA}, and e.g. Mel\'endez \& Barbuy~(\cite{melendez}), 
Mel\'endez et al.~(\cite{melendez2}), and 
Balachandran et al.~(\cite{balachandran}) for abundances from the IR OH lines. 
Other indicators such as the \ion{O}{i} triplet at 7774\,{\AA} and the UV OH 
lines give higher [O/Fe] values with an uprising trend when going to lower 
[Fe/H], see e.g. Israelian et al.~(\cite{israelian}) and 
Abia \& Rebolo~(\cite{abia}) for abundances from the \ion{O}{i} triplet, and 
Boesgaard et al.~(\cite{boesgaard}) and Israelian et al.~(\cite{israelian2}) 
for abundances from the UV OH lines. However, agreement between abundances 
from the UV OH lines and the forbidden lines has also been found 
(e.g. Bessel et al.~\cite{bessel}) and also between the \ion{O}{i} triplet 
and the forbidden lines  (e.g Carretta et al.~\cite{carretta}; 
Mishenina et al.~\cite{mishenina}). 

For stars more metal-rich than the halo (i.e. 
[Fe/H]\,$\gtrsim$\,$-1$ to $-0.5$) the observed trends are further complicated 
by the fact that there is a mixture of stars from different stellar 
populations, each with their own chemical history (see e.g. 
Gratton et al.~\cite{gratton2003}). The observed [O/Fe] trend for the halo 
stars generally extends to [Fe/H]\,$\sim$\,$-0.5$ with an overabundance of 
[O/Fe]\,$\sim$\,0.4\,--\,0.5. 
Nissen \& Schuster~(\cite{schuster}), however, found that the most metal-rich
halo stars to be less enhanced in [O/Fe] than the thick disk stars at the same
metallicities.
Quite good agreement between the different oxygen 
indicators are generally found at these metallicities (see works cited above). 
For the thick disk no (previous) study has included thick disk stars with 
[Fe/H]\,$\gtrsim$\,$-0.35$ when studying the [O/Fe] vs. [Fe/H] trend. Below 
this metallicity, thick disk stars are generally found to show an [O/Fe] trend 
similar to the halo trend, i.e. a constant value of 
[O/Fe]\,$\sim$\,0.4\,--\,0.5 (see e.g. Nissen et al.~\cite{nissen2002};
Tautvai\v sien\.e et al.~\cite{tautvaisiene}; 
Prochaska et al.~\cite{prochaska}; Gratton et al.~\cite{gratton}; 
Chen et al.~\cite{chen}). The thin disk stars present small to moderate 
overabundances of oxygen (0\,$\lesssim$\,[O/Fe]\,$\lesssim$\,0.2) for 
metallicities down to [Fe/H]\,$\approx$\,$-0.8$ with the largest values of 
[O/Fe] at the lower [Fe/H] values (see e.g. Reddy et al.~\cite{reddy}; 
Nissen et al.~\cite{nissen2002}; Edvardsson et al.~\cite{edvardsson}; 
Nissen \& Edvardsson~\cite{nissen1992}; Kj\ae rgaard et al.~\cite{kjaergaard}). 
The distinction between the thin and thick disk is however somewhat unclear. 
While e.g. Prochaska et al.~(\cite{prochaska}) found them to be distinct in 
terms of abundances, Chen et al.~(\cite{chen}) found the elemental abundance 
trends in the thin and thick disks to follow smoothly upon each other.

At super-solar metallicities ([Fe/H]\,$>$\,0) a constant value of 
[O/Fe]\,$\sim$\,0 was found by Nissen \& Edvardsson~(\cite{nissen1992}), 
and Nissen et al.~(\cite{nissen2002}).
Feltzing \& Gustafsson~(\cite{feltzing1998}) has a somewhat larger scatter in 
the [O/Fe] trend and also find a potential decline towards higher [Fe/H].
A decline in [O/Fe] is supported by the studies of 
Castro et al.~(\cite{castro}) and Chen et al.~(\cite{chen2003}),
but is at variance with other studies discussed above.

In the solar neighbourhood it is also possible to come across stars which might 
originate in the Galactic bulge or inner disk 
(Pomp\'eia et al.~\cite{pompeia}). They are generally old ($>$\,10\,Gyr)
and Pomp\'eia et al.~(\cite{pompeia}) found that they show an overabundance 
of oxygen when compared to their disk counterparts for metallicities 
ranging $-0.8$\,$\leq$\,[Fe/H]\,$\leq$\,+0.4.

In this paper we will determine the oxygen abundances for thin and thick disk 
stars with $-0.9$\,$\lesssim$\,[Fe/H]\,$\lesssim$\,+0.4. The stars have been 
selected on purely kinematical grounds (see Bensby et al.~\cite{bensby} for a 
full discussion of the selection process) in order to address two important 
questions:
1) The [O/Fe] trends at [Fe/H]\,$\lesssim$\,0 in the thin and thick disks; 
2) The [O/Fe] trend at super-solar metallicities.

Our oxygen abundances are based on the forbidden [\ion{O}{i}] line at 
6300\,{\AA}. In order to properly account for the Ni blends and the fact that 
the line generally is very weak we have obtained spectra of very high quality 
($R$\,$\gtrsim$\,215\,000 and $S/N$\,$\gtrsim$\,400). We have used the new 
$\log gf$-values for the Ni blends from Johansson et al.~(\cite{johansson}) who 
also showed that the Ni blend actually consists of two isotopic Ni components.

The paper is organized as follows. In Sect.~\ref{sec:observations} we describe 
the observations and the data reductions. In Sect.~\ref{sec:atomdata} we 
describe the atomic data used in the abundance determination, which in turn is 
described in Sect.~\ref{sec:abund_determ}. Section~\ref{sec:errors} 
deals with the errors in the resulting abundances. The abundance trends that 
we derive from the different indicators are presented in 
Sect.~\ref{sec:abund_trends}. In Sect.~\ref{sec:implications} we put our 
results into the contexts of Galactic chemical evolution.
We finally give a summary in Sect.~\ref{sec:summary}.

%==============================================================================
\section{Observations and data reductions}    \label{sec:observations}

%==============================================================================
\subsection{Stellar sample}  \label{sec:sample}

The stellar samples have been kinematically selected to contain stars with high 
probabilities of belonging to either the thin or the thick Galactic disks. 
These probabilities were derived under the assumption that the stellar space 
velocities $U_{\rm LSR}$, $V_{\rm LSR}$, and $W_{\rm LSR}$ in the thin and 
thick disks have Gaussian distributions. For each star it is then possible to 
calculate the probability that it kinematically belongs to either the thin or 
the thick disk. Since we are working with nearby stars we also need to take the 
local number density of the thin and thick disk stars into consideration. For 
the thick disk we used a density of 6\,\% in the solar neighbourhood and 
consequently 94\,\% for the thin disk. We then formed the TD/D ratio (i.e. the 
thick disk probability divided by the thin disk probability) for each star.
All thin disk stars in the sample (apart from three) have thin disk 
probabilities that 
are more than ten times larger than their thick disk probability 
(i.e. TD/D\,$<$\,0.1). Three thin disk stars (Hip 3170, 95447, 100412)
have higher TD/D ratios (but still approximately two 
times more probable of belonging to the thin disk than the thick disk). 
The majority of the thick disk stars are more than ten 
times more probable to belong to the thick disk than to the thin disk 
(i.e. TD/D\,$>$\,10). 
For a few thick disk stars we loosened this criterion somewhat in order to
increase the statistics, allowing thick disk stars with 1\,$<$\,TD/D\,$<$\,10 
in the sample. Since the two thick disk sub-samples 
(TD/D\,$>$\,10 and 1\,$<$\,TD/D\,$<$\,10) show exactly the same trends for 
other elements such as Na, Mg, Al, Si, Ca, Ti, Cr, Ni, and Zn 
(see Bensby et al.~\cite{bensby}), we will not distinguish between them in 
this study.

Apart from the three thin disk stars with slightly larger TD/D ratios,
two additional thin disk stars (Hip 5862 and Hip 15131), and one additional 
thick disk star (Hip 15510) the stellar sample is the same as in 
Bensby et al.~(\cite{bensby}) (i.e., 66 stars in common). The sample now 
consists of 22 thick disk stars and 50 thin disk stars
(see Table~\ref{tab:abundances}). 
The selection criteria and the stellar samples
are fully described in Bensby et al.~(\cite{bensby}). 
The additional 6 stars are described in a forthcoming
paper (Bensby et al., in prep.) where we moreover analyze a further 
30 thin and thick disk dwarf stars in the northern 
hemisphere, observed with the SOFIN spectrograph on the NOT telescope on 
La Palma.

%------------------------------------------------------------------------------
\begin{figure}
 \resizebox{\hsize}{!}{\includegraphics[bb=18 260 592 718,clip]{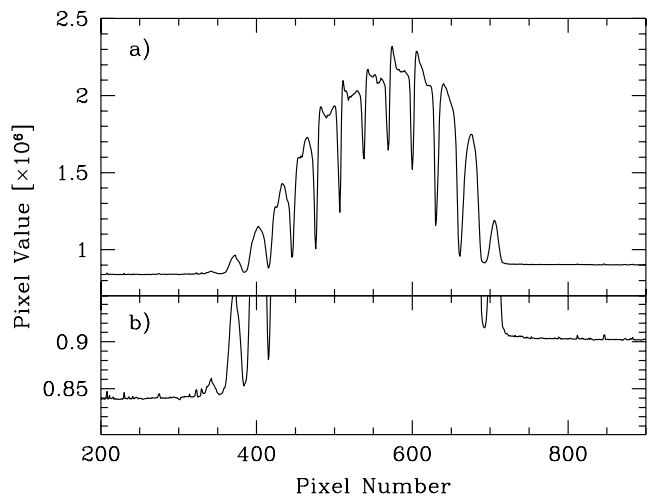}}
\caption{
	{\bf a)} The CES image slicer profile. 
	{\bf b)} A zoom in on the straylight pedestal beneath 
	the light pattern of the image slicer profile.
         }
\label{fig:pedestal}
\end{figure}
%------------------------------------------------------------------------------

%==============================================================================
\subsection{Observations} 

Observations were carried out with ESO's 3.6 m telescope on La Silla in Chile 
during two observing runs in September 2000 (4 nights, TB and SF as observers) 
and August/September 2001 (6 nights, TB as observer). We used the 
Coud\'e Echelle Spectrograph (CES) at its highest resolution, 
$R\! \simeq\! 215\,000$, centered on the forbidden oxygen line at 6300~{\AA} 
(hereafter denoted as [\ion{O}{i}]$_{6300}$). This setting gives a spectrum 
with a wavelength coverage of $\sim$\,40\,{\AA}. All spectra have a 
$S/N$ of at least 400. Long exposures times (given in 
Table~\ref{tab:abundances}) were usually split into three exposures. Since 
telluric lines are present in this wavelength region we also observed rapidly 
rotating B stars. These spectra were 
then used to divide out telluric lines from the object spectra.

Spectra covering the whole optical wavelength region from 3560\,{\AA} to 
9200\,{\AA} were also obtained with the FEROS spectrograph on the ESO\,1.5\,m 
telescope during the same observing runs. These spectra have 
$R$\,$\sim$\,$48\,000$ and $S/N$\,$\gtrsim$\,$150$ and provide Fe and Ni 
abundances as well as the atmospheric parameters for the stars 
(analysis presented in Bensby et al.~\cite{bensby}).

Spectra for the three additional stars (Hip 5862, 15131, 15510) 
were obtained with the UVES spectrograph on the VLT/UT2 
telescope Kueyen on Paranal in Chile in July 2002 (TB as observer). These 
spectra have $R$\,$\sim$\,110\,000 and $S/N$\,$\sim$\,350
(Bensby et al., in prep.).

%------------------------------------------------------------------------------
\begin{figure}
 \resizebox{\hsize}{!}{\includegraphics[bb=18 144 592 477,clip]{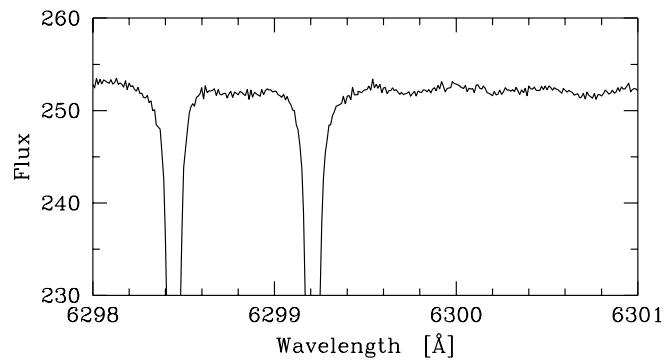}}
\caption{
        Example of the interference pattern that originates from reflections
        within the optical fiber. It is best seen in a featureless spectrum,
        in this case a fast rotating B star. The two telluric lines are the
        same as those in Fig.~\ref{fig:telluric}. The spectrum has been
        wavelength calibrated but not normalized and not corrected for the
        radial velocity of the star.
         }
\label{fig:bstar}
\end{figure}
%------------------------------------------------------------------------------

%==============================================================================
\subsection{Reduction of CES spectra}  \label{sec:reduction}

The reduction procedure from raw images to wavelength calibrated spectra was 
carried out with the MIDAS\footnote{ESO-MIDAS is the acronym for the European 
Southern Observatory Munich Image Data Analysis System which is developed and 
maintained by the European Southern Observatory.} software. The reductions are 
straightforward using standard routines concerning cosmic ray filtering, 
background subtraction, flatfield division, wavelength calibration, and 
continuum fitting. All exposures for a single star were reduced individually 
and then the extracted spectra were co-added before the setting of the 
continuum. For those stars that had a telluric absorption line coinciding with 
the [\ion{O}{i}]$_{6300}$ line we divided the spectra with a featureless 
spectrum from a fast rotating B star. Finally the co-added spectra were shifted 
to rest wavelengths by measuring the doppler shift of eight evenly distributed 
\ion{Fe}{i} lines that have accurate wavelengths from 
Nave et al.~(\cite{nave}). There are, however, some properties of the CES 
spectra that are worth discussing in further detail.

\paragraph {Straylight pedestal:}
The background level is different on the two sides of the spectrum in the 
direction perpendicular to the dispersion direction 
(see Fig.~\ref{fig:pedestal}). This is most likely diffuse light produced by
the image slicer itself (K\"urster~\cite{kurster}). In the wavelength 
direction this background is slightly non-uniform. Actually it is a 
low-resolution version of the stellar spectrum that for all practical
purposes can be treated as featureless 
(K\"urster~\cite{kurster}). The [\ion{O}{i}]$_{6300}$ line profile 
from individual slices were checked to see if this straylight pedestal had any 
influence on the line profile and strength from different slices. 
We especially checked for 
variations in the line strength (equivalent width) of the 
[\ion{O}{i}]$_{6300}$ line. No discernible effects (apart from differences in 
the $S/N$) were, however, found between the slices.

%-------------------------------------------------------------------------------
\begin{figure}
 \resizebox{\hsize}{!}{\includegraphics{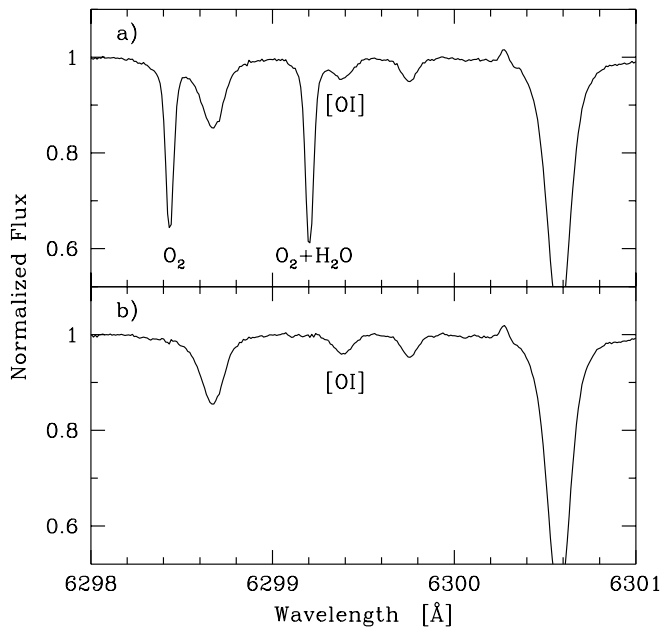}}
\caption{
        Example of the removal of lines originating in Earth's atmosphere. The
        telluric lines, O$_{2}$ at 6298.5 {\AA}, and O$_{2}$+H$_{2}$O at
        6299.2\,{\AA} are indicated as well as the [\ion{O}{i}]$_{6300}$ line.
        {\bf a)} Spectrum of Hip 75181 prior to removal of telluric lines
        and {\bf b)} after removal of telluric lines by division with a
        spectrum of a fast rotating B star. The spectrum has been wavelength
        calibrated but not corrected for the radial velocity of the star. Note
        the telluric oxygen emission line at $\sim\!6300.25$\,{\AA}.
         }
\label{fig:telluric}
\end{figure}
%------------------------------------------------------------------------------

\paragraph {Interference pattern:}
A regular sinusoidal pattern can also be seen in the continuum of a featureless 
CES spectrum (see Fig.~\ref{fig:bstar}). Most likely this has its origin from 
reflections within the optical fiber that feeds the spectrograph 
(K\"urster~\cite{kurster}). For low metallicity stars with a low density of 
spectral lines it is possible to perform a Fourier analysis of the pattern and 
remove it. However, most of our stars are rich in spectral lines which 
makes a removal nearly impossible. Our B star spectra indicate that the 
amplitude of this pattern is very small, typically 0.2\% of the continuum flux 
(see Fig.~\ref{fig:bstar}). Unless spectra with $S/N$\,$>$\,500 are striven 
for, it is therefore safe to ignore this feature in the reduction process.

\paragraph {Telluric absorption lines:}
Telluric lines were divided out using spectra of fast rotating B stars with the
IRAF\footnote{IRAF is distributed by National Optical Astronomy Observatories,
operated by the Association of Universities for Research in Astronomy, Inc.,
under contract with the National Science Foundation, USA.} task {\sc telluric}
which is part of the {\sc noao} package. The procedure takes care of small
possible shifts in the dispersion zero-points as well as of the variations in
the abundance of the telluric species by shifting and scaling the B star
spectrum. An example of the removal of the telluric absorption lines is given
in Fig.~\ref{fig:telluric}. Of the 72 stars in our sample there were eight
stars (Hip 3086, 3185, 3704, 15510, 75181, 82588, 88622, and 95447) that had a
telluric absorption line that blended the [\ion{O}{i}]$_{6300}$ line. Only for
these eight stars did we divide out the telluric lines.

\paragraph {Telluric emission lines:}
Unfortunately there were also occasions when the telluric oxygen emission line
(see Fig.~\ref{fig:telluric}) coincides with the stellar [\ion{O}{i}]$_{6300}$ 
line. Since the presence and strength of
this feature vary strongly with time it is very difficult to remove.
A possible way to remove it would be to observe a part of the sky next to the
object immediately after the observation of the star, and then use this sky
spectrum to divide out the sky emission. We did, however, not obtain
such sky spectra. For seven stars (Hip 3170, 9085, 12611, 22263, 86731, 90485,
83229) we were unable to determine the oxygen abundance based on the
[\ion{O}{i}]$_{6300}$ line due to this telluric emission feature.

%------------------------------------------------------------------------------
\begin{table*}[ht]
\caption{
        The 72 program stars. $T_{\rm eff}$, $\log g$, $\xi_{\rm t}$, 
	[Fe/H], and [Ni/H] are taken from Bensby et al.~(\cite{bensby}). 
	All abundances have been normalized to the Sun. Both NLTE corrected, 
	according to the prescription in Gratton et al.~(\cite{gratton}), and 
	un-corrected abundances are given for the triplet lines. For oxygen 
	abundances from the individual triplet lines see 
	Table~\ref{tab:o7774abundances}. The last three columns give the 
	spectrographs and exposure times used: C$=$CES (ESO\,3.6\,m), 
	F$=$FEROS (ESO\,1.5\,m), and U$=$UVES (ESO VLT/UT2). 
        }
\centering \scriptsize
\begin{tabular}{rrrccccrrrrrlll}
\hline \hline\noalign{\smallskip}
          \multicolumn{3}{c}{Identifications}
        & $T_{\rm eff}$
        & $\log g$
        & $\xi_{\rm t}$
        & $v_{\rm rot-macro}$
        & [Fe/H]
        & [Ni/H]
	& [O/H]
	& [O/H]
	& [O/H]
        & \multicolumn{3}{c}{Spectrograph (EXPTIME)} \\
\noalign{\smallskip}
          Hip
        & HD
        & HR
        & [K]
        & [cgs]
        & [km\,s$^{-1}$]
        &  [km\,s$^{-1}$]
        &
        &
        & 6300     
        & 6363     
        & 7774     
        & \multicolumn{3}{c}{[min]}      \\
\noalign{\smallskip}
\hline
\noalign{\smallskip}
\multicolumn{15}{c}{\bf THIN DISK STARS} \\
\noalign{\smallskip}
    Sun   &        &      & 5777 & 4.44 & 0.85 & 1.74 &     0.00  &    0.00  &   0.00  &   0.00    &  $ 0.00$  &   C      &  F       &    U            \\
    3142  &   3735 &  170 & 6100 & 4.07 & 1.50 & 2.22 &   $-0.45$ &  $-0.47$ & $-0.25$ &           &  $-0.19$  &   C (80) &  F (9)   &                 \\
    3170  &   3823 &  176 & 5970 & 4.11 & 1.40 & 2.13 &   $-0.34$ &  $-0.37$ &         &           &  $-0.06$  &   C (30) &  F (10)  &                 \\
    5862  &   7570 &  370 & 6100 & 4.26 & 1.10 & 2.80 &   $ 0.17$ &  $ 0.16$ & $ 0.06$ &           &           &   C (20) &          & U (0.3)         \\
    7276  &   9562 &  448 & 5930 & 3.99 & 1.35 & 2.61 &   $ 0.20$ &  $ 0.24$ & $ 0.14$ &           &  $ 0.12$  &   C (30) &  F (4)   &                 \\
    9085  &  12042 &  573 & 6200 & 4.25 & 1.30 & 3.19 &   $-0.31$ &  $-0.37$ &         &           &  $-0.05$  &   C (28) &  F (6.3) & U (1)           \\
   10798  &  14412 &  683 & 5350 & 4.57 & 0.20 & 1.51 &   $-0.47$ &  $-0.48$ & $-0.30$ &           &  $-0.37$  &   C (45) &  F (7)   &                 \\
   12186  &  16417 &  772 & 5800 & 4.04 & 1.20 & 1.94 &   $ 0.14$ &  $ 0.14$ & $ 0.08$ &           &  $ 0.13$  &   C (30) &  F (4.3) &                 \\
   12611  &  17006 &  807 & 5250 & 3.66 & 1.35 & 2.51 &   $ 0.26$ &  $ 0.33$ &         & $ 0.14$   &  $ 0.34$  &   C (35) &  F (6)   &                 \\
   12653  &  17051 &  810 & 6150 & 4.37 & 1.25 & 3.43 &   $ 0.14$ &  $ 0.12$ & $ 0.04$ & $ 0.06$   &  $ 0.20$  &   C (16) &  F (3)   &                 \\
   14954  &  19994 &  962 & 6240 & 4.10 & 1.60 & 4.40 &   $ 0.19$ &  $ 0.22$ & $ 0.11$ &           &  $ 0.18$  &   C (40) &  F (3)   &                 \\
   15131  &  20407 &      & 5834 & 4.35 & 1.00 & 1.58 &   $-0.52$ &  $-0.52$ & $-0.26$ &           &           &          &          & U (1.5)         \\
   17378  &  23249 & 1136 & 5020 & 3.73 & 0.80 & 1.80 &   $ 0.24$ &  $ 0.31$ & $ 0.13$ & $ 0.05$   &  $ 0.08$  &   C (12) &  F (0.5) &                 \\
   22263  &  30495 & 1532 & 5850 & 4.50 & 0.95 & 2.40 &   $ 0.05$ &  $-0.01$ &         &           &  $ 0.06$  &   C (21) &  F (3.3) &                 \\
   22325  &  30606 & 1538 & 6250 & 3.91 & 1.80 & 3.90 &   $ 0.06$ &  $ 0.02$ & $ 0.01$ &           &  $ 0.20$  &   C (108)&  F (4)   &                 \\
   23555  &  32820 & 1651 & 6300 & 4.29 & 1.50 & 4.58 &   $ 0.13$ &  $ 0.11$ & $ 0.06$ &           &  $ 0.21$  &   C (34) &  F (6.7) &                 \\
   23941  &  33256 & 1673 & 6427 & 4.04 & 1.90 & 5.20 &   $-0.30$ &  $-0.36$ &         &           &  $-0.06$  &          &  F (4.7) &                 \\
   24829  &  35072 & 1767 & 6360 & 3.93 & 1.70 & 2.37 &   $ 0.06$ &  $ 0.01$ & $ 0.03$ &           &  $ 0.16$  &   C (20) &  F (3.3) &                 \\
   29271  &  43834 & 2261 & 5550 & 4.38 & 0.80 & 1.68 &   $ 0.10$ &  $ 0.16$ & $ 0.02$ &           &  $ 0.15$  &   C (20) &  F (3)   &                 \\
   30480  &  45701 & 2354 & 5890 & 4.15 & 1.20 & 2.11 &   $ 0.19$ &  $ 0.22$ & $ 0.17$ & $ 0.10$   &  $ 0.20$  &   C (50) &  F (8.3) &                 \\
   30503  &  45184 & 2318 & 5820 & 4.37 & 0.90 & 2.24 &   $ 0.04$ &  $ 0.02$ & $ 0.02$ & $ 0.00$   &  $ 0.10$  &   C (41) &  F (8.3) &                 \\
   72673  & 130551 &      & 6350 & 4.18 & 1.60 & 2.30 &   $-0.62$ &  $-0.60$ & $-0.30$ &           &  $-0.34$  &   C (80) &  F (15)  &                 \\
   78955  & 144585 & 5996 & 5880 & 4.22 & 1.12 & 2.32 &   $ 0.33$ &  $ 0.40$ & $ 0.17$ & $ 0.17$   &  $ 0.23$  &   C (60) &  F (6)   &                 \\
   80337  & 147513 & 6094 & 5880 & 4.49 & 1.10 & 2.09 &   $ 0.03$ &  $-0.04$ & $ 0.03$ &           &  $-0.04$  &   C (20) &  F (3)   &                 \\
   80686  & 147584 & 6098 & 6090 & 4.45 & 1.01 & 2.28 &   $-0.06$ &  $-0.13$ & $-0.05$ & $-0.09$   &  $ 0.01$  &   C (20) &  F (2)   &                 \\
   81520  & 149612 &      & 5680 & 4.53 & 0.65 & 1.70 &   $-0.48$ &  $-0.49$ &         &           &  $-0.24$  &          &  F (11)  &                 \\
   83601  & 154417 & 6349 & 6167 & 4.48 & 1.21 & 3.27 &   $ 0.09$ &  $ 0.00$ & $ 0.07$ &           &  $ 0.04$  &   C (34) &  F (5)   &                 \\
   84551  & 156098 & 6409 & 6475 & 3.79 & 2.00 & 3.98 &   $ 0.12$ &  $ 0.08$ & $ 0.12$ &           &  $ 0.28$  &   C (18) &  F (3.3) &                 \\
   84636  & 156365 &      & 5820 & 3.91 & 1.30 & 2.68 &   $ 0.23$ &  $ 0.32$ & $ 0.15$ &           &  $ 0.28$  &   C (55) &  F (9)   &                 \\
   85007  & 157466 &      & 6050 & 4.37 & 1.10 & 2.43 &   $-0.39$ &  $-0.45$ & $-0.26$ &           &  $-0.27$  &   C (75) &  F (10)  &                 \\
   85042  & 157347 & 6465 & 5720 & 4.40 & 1.00 & 1.65 &   $ 0.03$ &  $ 0.06$ & $ 0.00$ & $ 0.06$   &  $-0.05$  &   C (40) &  F (6.7) &                 \\
   86731  & 161239 & 6608 & 5840 & 3.79 & 1.43 & 3.37 &   $ 0.25$ &  $ 0.33$ &         &           &  $ 0.26$  &   C (39) &  F (5)   &                 \\
   86796  & 160691 & 6585 & 5800 & 4.30 & 1.05 & 2.20 &   $ 0.32$ &  $ 0.41$ & $ 0.14$ &           &  $ 0.24$  &   C (27) &  F (2.5) &                 \\
   87523  & 162396 & 6649 & 6070 & 4.07 & 1.36 & 2.17 &   $-0.40$ &  $-0.39$ & $-0.31$ & $-0.31$   &  $-0.23$  &   C (35) &  F (6)   &                 \\
   90485  & 169830 & 6907 & 6339 & 4.05 & 1.55 & 2.59 &   $ 0.12$ &  $ 0.11$ &         & $ 0.06$   &  $ 0.25$  &   C (31) &  F (5)   &                 \\
   91438  & 172051 & 6998 & 5580 & 4.42 & 0.55 & 1.62 &   $-0.24$ &  $-0.26$ & $-0.18$ &           &  $-0.23$  &   C (45) &  F (4)   &                 \\
   94645  & 179949 & 7291 & 6200 & 4.35 & 1.20 & 3.90 &   $ 0.16$ &  $ 0.11$ & $ 0.06$ & $ 0.11$   &  $ 0.23$  &   C (40) &  F (6.5) &                 \\
   95447  & 182572 & 7373 & 5600 & 4.13 & 1.10 & 2.14 &   $ 0.37$ &  $ 0.47$ & $ 0.20$ & $ 0.23$   &  $ 0.38$  &   C (18) &  F (2.5) &                 \\
   96536  & 184985 & 7454 & 6397 & 4.06 & 1.65 & 2.72 &   $ 0.03$ &  $ 0.00$ & $ 0.05$ &           &  $ 0.20$  &   C (21) &  F (3.3) &                 \\
   98785  & 190009 & 7658 & 6430 & 3.97 & 1.90 & 4.25 &   $ 0.03$ &  $-0.03$ & $ 0.11$ & $ 0.01$   &  $ 0.18$  &   C (60) &  F (7.5) &                 \\
   99240  & 190248 & 7665 & 5585 & 4.26 & 0.98 & 1.72 &   $ 0.37$ &  $ 0.44$ & $ 0.19$ & $ 0.17$   &  $ 0.32$  &   C (4)  &  F (1)   &                 \\
  100412  & 193307 & 7766 & 5960 & 4.06 & 1.20 & 2.18 &   $-0.32$ &  $-0.36$ & $-0.18$ &           &  $-0.03$  &   C (45) &  F (6)   &                 \\
  102264  & 197214 &      & 5570 & 4.37 & 0.60 & 1.81 &   $-0.22$ &  $-0.26$ & $-0.12$ &           &  $-0.15$  &   C (80) &  F (11)  &                 \\
  103682  & 199960 & 8041 & 5940 & 4.26 & 1.25 & 2.34 &   $ 0.27$ &  $ 0.35$ & $ 0.14$ & $ 0.12$   &  $ 0.21$  &   C (70) &  F (6.5) &                 \\
  105858  & 203608 & 8181 & 6067 & 4.27 & 1.17 & 2.04 &   $-0.73$ &  $-0.74$ & $-0.46$ &           &  $-0.34$  &   C (30) &  F (1)   &                 \\
  109378  & 210277 &      & 5500 & 4.30 & 0.78 & 1.71 &   $ 0.22$ &  $ 0.26$ & $ 0.16$ & $ 0.16$   &  $ 0.24$  &   C (60) &  F (8)   &                 \\
  110341  & 211976 & 8514 & 6500 & 4.29 & 1.70 & 3.63 &   $-0.17$ &  $-0.24$ & $-0.11$ & $-0.12$   &  $-0.01$  &   C (45) &  F (6)   &                 \\
  113137  & 216437 & 8701 & 5800 & 4.10 & 1.16 & 2.17 &   $ 0.22$ &  $ 0.27$ & $ 0.13$ & $ 0.12$   &  $ 0.27$  &   C (35) &  F (5.5) &                 \\
  113357  & 217014 & 8729 & 5789 & 4.34 & 1.00 & 1.91 &   $ 0.20$ &  $ 0.26$ & $ 0.10$ & $ 0.24$   &  $ 0.11$  &   C (21) &  F (3.2) &                 \\
  113421  & 217107 & 8734 & 5620 & 4.29 & 0.97 & 1.75 &   $ 0.35$ &  $ 0.42$ & $ 0.16$ &           &  $ 0.27$  &   C (45) &  F (6)   &                 \\
  117880  & 224022 & 9046 & 6100 & 4.21 & 1.30 & 3.16 &   $ 0.12$ &  $ 0.14$ & $ 0.05$ &           &  $ 0.16$  &   C (60) &  F (5.5) &                 \\
\noalign{\smallskip}
\hline
\end{tabular}
\label{tab:abundances}
\end{table*}
%------------------------------------------------------------------------------
\setcounter{table}{0}
\begin{table*}[ht]
\caption{
	{\it continued}
        }
\centering \scriptsize
\begin{tabular}{rrrccccrrrrrrlll}
\hline \hline\noalign{\smallskip}
          \multicolumn{3}{c}{Identifications}
        & $T_{\rm eff}$
        & $\log g$
        & $\xi_{\rm t}$
        & $v_{\rm rot-macro}$
        & [Fe/H]
        & [Ni/H]
	& [O/H]
	& [O/H]
	& [O/H]
        & \multicolumn{3}{c}{Spectrograph (EXPTIME)} \\
\noalign{\smallskip}
          Hip
        & HD
        & HR
        & [K]
        & [cgs]
        & [km\,s$^{-1}$]
        &  [km\,s$^{-1}$]
        &
        & 
        & 6300
        & 6363
        & 7774     
        & \multicolumn{3}{c}{[min]}      \\
\noalign{\smallskip}
\hline
\noalign{\smallskip}
\multicolumn{15}{c}{\bf THICK DISK STARS}   \\
\noalign{\smallskip}
    3086  &   3628 &      & 5840 & 4.15 & 1.15 & 1.90 &   $-0.11$ &  $-0.12$ & $ 0.09$ &           &  $-0.04$  &   C (100)&  F (10)  &                 \\
    3185  &   3795 &  173 & 5320 & 3.78 & 0.70 & 1.64 &   $-0.59$ &  $-0.57$ & $-0.14$ & $-0.21$   &  $-0.11$  &   C (45) &  F (6.7) &                 \\
    3497  &   4308 &      & 5636 & 4.30 & 0.80 & 1.76 &   $-0.33$ &  $-0.31$ & $-0.05$ & $-0.10$   &  $ 0.03$  &   C (45) &  F (10)  &                 \\
    3704  &   4597 &      & 6040 & 4.30 & 1.08 & 2.10 &   $-0.38$ &  $-0.40$ & $-0.08$ &           &  $-0.25$  &   C (180)&  F (10)  &                 \\
    5315  &   6734 &      & 5030 & 3.46 & 0.86 & 1.66 &   $-0.42$ &  $-0.36$ & $ 0.00$ & $-0.01$   &  $ 0.03$  &   C (45) &  F (10)  &                 \\
   14086  &  18907 &  914 & 5110 & 3.51 & 0.87 & 1.64 &   $-0.59$ &  $-0.53$ & $-0.16$ &           &  $-0.20$  &   C (42) &  F (5)   &                 \\
   15510  &  20794 & 1008 & 5480 & 4.43 & 0.75 & 1.28 &   $-0.41$ &  $-0.36$ & $-0.01$ & $-0.10$   &           &   C (50) &          & U (0.2)         \\
   17147  &  22879 &      & 5920 & 4.33 & 1.20 & 1.74 &   $-0.84$ &  $-0.83$ & $-0.32$ &           &  $-0.32$  &   C (107)&  F (10)  &                 \\
   75181  & 136352 & 5699 & 5650 & 4.30 & 0.78 & 1.69 &   $-0.34$ &  $-0.34$ & $ 0.00$ &           &  $ 0.01$  &   C (22) &  F (5)   &                 \\
   79137  & 145148 & 6014 & 4900 & 3.62 & 0.60 & 1.53 &   $ 0.30$ &  $ 0.39$ & $ 0.20$ & $ 0.13$   &  $ 0.36$  &   C (28) &  F (4)   &                 \\
   82588  & 152391 &      & 5470 & 4.55 & 0.90 & 2.53 &   $-0.02$ &  $-0.06$ & $ 0.01$ & $ 0.00$   &  $-0.01$  &   C (54) &  F (8)   &                 \\
   83229  & 153075 &      & 5770 & 4.17 & 0.97 & 1.52 &   $-0.57$ &  $-0.51$ &         & $-0.23$   &  $-0.10$  &          &  F (11)  &                 \\
   84905  & 157089 &      & 5830 & 4.06 & 1.20 & 1.98 &   $-0.57$ &  $-0.56$ & $-0.15$ &           &  $-0.17$  &   C (75) &  F (11)  &                 \\
   88622  & 165401 &      & 5720 & 4.35 & 0.80 & 2.10 &   $-0.46$ &  $-0.45$ & $-0.05$ & $-0.09$   &  $ 0.02$  &   C (60) &  F (9)   &                 \\
   96124  & 183877 &      & 5590 & 4.37 & 0.78 & 1.78 &   $-0.20$ &  $-0.17$ & $ 0.08$ & $ 0.08$   &  $ 0.07$  &   C (80) &  F (13)  &                 \\
   98767  & 190360 & 7670 & 5490 & 4.23 & 0.66 & 1.98 &   $ 0.25$ &  $ 0.31$ & $ 0.21$ & $ 0.13$   &  $ 0.38$  &   C (25) &  F (5)   &                 \\
  103458  & 199288 &      & 5780 & 4.30 & 0.90 & 1.69 &   $-0.65$ &  $-0.64$ & $-0.29$ &           &  $-0.29$  &   C (60) &  F (8)   &                 \\
  108736  & 208988 &      & 5890 & 4.24 & 1.05 & 1.97 &   $-0.38$ &  $-0.35$ & $-0.01$ & $-0.01$   &  $ 0.06$  &   C (80) &  F (14)  &                 \\
  109450  & 210483 &      & 5830 & 4.18 & 1.10 & 1.43 &   $-0.13$ &  $-0.15$ & $ 0.05$ & $ 0.08$   &  $ 0.06$  &          &  F (10)  &                 \\
  109821  & 210918 & 8477 & 5800 & 4.29 & 1.05 & 1.72 &   $-0.08$ &  $-0.10$ & $ 0.10$ &           &  $ 0.01$  &   C (45) &  F (6.7) &                 \\
  110512  & 212231 &      & 5770 & 4.15 & 1.05 & 2.12 &   $-0.30$ &  $-0.28$ & $ 0.00$ &           &  $ 0.07$  &   C (180)&  F (10)  &                 \\
  118115  & 224383 &      & 5800 & 4.30 & 1.00 & 1.20 &   $-0.01$ &  $-0.04$ & $ 0.04$ &           &  $-0.01$  &          &  F (10)  &                 \\
\noalign{\smallskip}
\hline
\end{tabular}
\end{table*}
%------------------------------------------------------------------------------

%==============================================================================
\section{Atomic data}   \label{sec:atomdata}

The oscillator strength of the weak Ni line at 6300.35\,{\AA} has recently
been measured in the laboratory by Johansson et al.~(\cite{johansson}). They
showed that it actually contains two isotopic components,
$\rm \lambda(^{58}Ni)$\,$=$\,6300.335\,{\AA} and
$\rm \lambda(^{60}Ni)$\,$=$\,6300.355\,{\AA}. Both components have
$\log gf$\,$=$\,$-2.11$. Assuming a solar isotopic ratio of 0.38 for the
abundances of $\rm ^{60}Ni$ and $\rm ^{58}Ni$ and that these two isotopes make
up 94.4\% of the total Ni abundance (CRC~\cite{lide}), the weighted
$\log gf$-values for these isotopes are
$\log gf (^{58}{\rm Ni})$\,$=$\,$-2.275$  and
$\log gf (^{60}{\rm Ni})$\,$=$\,$-2.695$.

The $\log gf$-values for the oxygen lines have been taken from the VALD 
database (Piskunov et al.~\cite{piskunov}; Kupka et al.~\cite{kupka};
Ryabchikova et al.~\cite{ryabchikova}). The source for the [\ion{O}{i}] lines
is the compilation of Wiese et al.~(\cite{wiese}) who based the transition
probabilities on theoretical calculations from
Stoffregen \& Derblom~(\cite{stoffregen}), Omholt~(\cite{omholt}),
Naqvi~(\cite{naqvi}), and Yamanouchi \& Horie~(\cite{yamanouchi}). For the
\ion{O}{i} triplet lines the original sources are
Bi\'emont \& Zeippen~(\cite{biemont}) and Hibbert et al.~(\cite{hibbert}).

The broadening of atomic lines by radiation damping was considered in the
determination of the abundances and the damping constants ($\gamma_{\rm rad}$)
for the different lines were collected from the VALD database.

Collisional broadening, or Van der Waals broadening, was also considered. The
width cross-sections are taken from Anstee \& O'Mara~(\cite{anstee}),
Barklem \& O'Mara~(\cite{barklem}, \cite{barklem3}), and
Barklem et al.~(\cite{barklem2}, \cite{barklem4}). Lines present in these works
have been marked by an ``S" in Table~\ref{tab:linelist}. For spectral lines not
present in these works (marked by a ``U" in Table~\ref{tab:linelist}) we apply
the correction term ($\delta \gamma_{6}$) to the classical Uns\"old
approximation of the Van der Waals damping. Following
M\"ackle et al.~(\cite{mackle}) this parameter was set to 2.5 for the lines
considered here.

The atomic data are summarized in Table~\ref{tab:linelist}.

%-------------------------------------------------------------------------------
\begin{table}[h]
\caption{
        Atomic line data for the different oxygen lines and the blending Ni
        lines. Column~1 gives the element, col.~2 the wavelength, col.~3 
	the lower excitation potential, and col.~4 the correction factor to the
        classical Uns\"old damping constant. Column~5 indicates if the
	broadening by collisions have been taken from 
	Anstee \& O'Mara~(\cite{anstee}), Barklem \& O'Mara~(\cite{barklem}, 
	\cite{barklem3}), and Barklem et al.~(\cite{barklem2}, \cite{barklem4}) 
	(indicated by an ``S") instead of the classical Uns\"old broadening 
	(indicated by an ``U"). Column~6 gives the radiation
        damping constant and col.~7 the $\log gf$-values (references are given
	in Sect.~\ref{sec:atomdata}).
        }
\centering
\label{tab:linelist}
\begin{tabular}{ccccccr}
\hline   \hline\noalign{\smallskip}
El. &      $\lambda$ &  $\chi_{\rm l}$ & $\delta \gamma_{6}$ & DMP  & $\gamma_{\rm rad}$   & $\log gf$ \\
    &        (\AA )  &    (eV)         &                     &      &   (s$^{-1}$)         &           \\
\noalign{\smallskip}
\hline\noalign{\smallskip}
$\rm [\ion{O}{i}]$ &     6300.304 &   0.00 &  2.50 & U &  1.0e+05 & $ -9.819$                        \\
$\rm [\ion{O}{i}]$ &     6363.776 &   0.02 &  2.50 & U &  1.0e+05 & $-10.303$                        \\
       \ion{O}{i}  &     7771.944 &   9.14 &  2.50 & S &  4.8e+07 &     0.37~~                       \\
       \ion{O}{i}  &     7774.166 &   9.14 &  2.50 & S &  4.7e+07 &     0.22~~                       \\
       \ion{O}{i}  &     7775.388 &   9.14 &  2.50 & S &  4.5e+07 &     0.00~~                       \\
\noalign{\smallskip}
  $\rm ^{58}Ni$    &     6300.335 &   4.27 &  2.50 & U &  2.7e+08 & $ -2.275$                        \\
  $\rm ^{60}Ni$    &     6300.355 &   4.27 &  2.50 & U &  2.7e+08 & $ -2.695$                        \\
\noalign{\smallskip}
\hline
\end{tabular}
\end{table}
%------------------------------------------------------------------------------

%==============================================================================
\section{Abundance determination} \label{sec:abund_determ}

Oxygen abundances have been determined by line synthesis of the
[\ion{O}{i}]$_{6300}$ line (Sect.~\ref{sec:syre6300}), by equivalent width
measurements of the [\ion{O}{i}]$_{6363}$ line (Sect.~\ref{sec:syre6363}) and
the \ion{O}{i} triplet lines at $\sim$\,7774\,{\AA} (Sect.~\ref{sec:syre7774}).
The actual generation of synthetic spectra was done with the Uppsala
{\sc Spectrum} synthesis program and for the calculation of elemental
abundances from measured equivalent widths we used the Uppsala {\sc Eqwidth}
abundance program. The
stellar model atmospheres are standard 1-D LTE and were calculated with the
Uppsala MARCS code (Gustafsson et al.~\cite{gustafsson}; 
Edvardsson et al.~\cite{edvardsson}; Asplund et al.~\cite{asplund2}).

%==============================================================================
\subsection{Stellar atmosphere parameters}

Stellar model atmosphere parameters for 66 of the 72 stars in the sample were
derived in Bensby et al.~(\cite{bensby}) using $\sim$\,150 \ion{Fe}{i} lines
measured in the FEROS spectra. 
The parameters for the additional 6 stars that are included here have been 
determined in the same way (using UVES spectra for 3 of them) and will
be presented in Bensby et al.~(in prep.) (together with the SOFIN stars, 
see Sect.~\ref{sec:observations}). In summary, 
the tuning of the model
atmospheres includes the following steps: effective temperatures were
determined by requiring \ion{Fe}{i} lines with different lower excitation
potentials to produce equal abundances; surface gravities were determined using
Hipparcos parallaxes and stellar masses; microturbulences were determined by
forcing the \ion{Fe}{i} lines to give the same abundances regardless of line
strength. The atmospheric parameters
([Fe/H], $T_{\rm eff}$, $\log g$, and $\xi_{\rm t}$) for the whole sample are
given together with the derived oxygen abundances in
Table~\ref{tab:abundances}.

%==============================================================================
\subsection{Macroscopic line broadening}  \label{sec:rotation}

%-------------------------------------------------------------------------------
\begin{figure}
\centering
\resizebox{\hsize}{!}{\includegraphics[bb=18 235 565 735,clip]{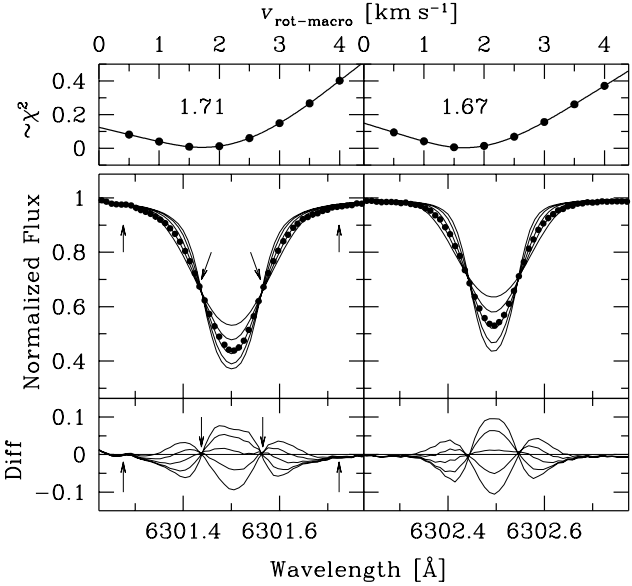}}
\caption{
        Determination of $v_{\rm rot-macro}$
        for Hip 75181. The two \ion{Fe}{i} lines at 6301.501\,{\AA}
        (left panels) and 6302.494\,{\AA} (right panels) have been used. The
        top panels show the un-normalized $\chi^2$-values for different
	values of
        $v_{\rm rot-macro}$ in steps of 0.5\,km\,s$^{-1}$ and a spline
        function connecting them. The $v_{\rm rot-macro}$ for which the
        $\chi^2$-functions are minimized are also indicated. The middle panels
        show the observed spectra (solid circles) and synthetic spectra for
        five different values of $v_{\rm rot-macro}$
        (0.5,\,1,\,1.5,\,2,\,2.5\,km\,s$^{-1}$). The bottom 
        panels show the difference between the observed and the synthetic 
        spectra in the middle panels. Up-arrows indicate where
        the residuals converge, and down-arrows where they
        intersect. 
         }
\label{fig:rotation}
\end{figure}
%------------------------------------------------------------------------------

Apart from line broadening due to atomic properties the line profile is 
affected by the instrument profile, 
the line-of-sight component of the stellar rotation 
($v_{\rm rot}\cdot \sin i$), and the macroturbulence. 
The instrument profile broadening is set by the resolution of the spectra 
($R$). To determine the contributions from the other two broadening mechanisms 
we used the two strong \ion{Fe}{i} lines at 6301.501\,{\AA} and 6302.494\,{\AA} 
(the former can be 
seen to the right in Fig.~\ref{fig:telluric}). So, in order to broaden the 
spectral lines we first convolved the synthetic spectra with a Gaussian 
profile in order to produce the instrumental broadening and then with an 
elliptical profile to produce the combined effect of the stellar rotation and 
macroturbulence. Since we have chosen to include the macroturbulence 
broadening into the elliptical broadening of the rotation we refer to this
profile as $v_{\rm rot-macro}$.

In Fig.~\ref{fig:rotation} we illustrate the process for Hip 75181. The first
step was to tune the $\log gf$-values of the two \ion{Fe}{i} lines in order to
make the shape of the synthetic line profiles similar to the observed lines.
This was done since the Fe abundances that were used in the models have been
determined from $\sim$\,150 other, generally weaker, \ion{Fe}{i} lines
(see Bensby et al.~\cite{bensby}). We created 10 synthetic spectra for each
star with $v_{\rm rot-macro}$ ranging from 0.5 to 5\,km\,s$^{-1}$ in steps of
0.5\,km\,s$^{-1}$. 
For each $v_{\rm rot-macro}$ we estimated the (un-normalized)
$\chi^2$-value (i.e. the added sum of squares of the difference between
observed and synthetic spectra in a certain wavelength interval).

When satisfactory $\log gf$-values have been reached the intersection points
(indicated by down-arrows in Fig.~\ref{fig:rotation}) of the
residuals between observed and synthetic spectra should lie on a
vertical position equivalent to the level farther from the line core (i.e.
where one approaches the continuum) and the residuals for the
different $v_{\rm rot-macro}$ converge (indicated by up-arrows in
Fig.~\ref{fig:rotation}).

Next, the $\chi^{2}$-functions between observed and synthetic spectra were
minimized in order to find the $v_{\rm rot-macro}$ (see top panels in
Fig.~\ref{fig:rotation}), using the new $\log gf$-values. We adopted the 
average $v_{\rm rot-macro}$ as derived
from the two Fe lines whenever possible (sometimes one of the two was affected
by telluric lines). These values are listed in column 7 in
Table~\ref{tab:abundances}.

%==============================================================================
\subsection{Solar analysis}  \label{sec:solaranalysis}

Our solar analysis is based on integrated solar light spectra that were
obtained by observing the Moon, Jupiter's moon Ganymede, and the afternoon sky
with CES. The solar UVES spectrum is the Moon spectrum that can be obtained
from ESO's 
web-pages\footnote{\tt http://www.eso.org/observing/dfo/quality/UVES/ \\pipeline/solar\_spectrum.html}
and the solar FEROS spectra were obtained by observing the afternoon sky.
The atmospheric parameters for our solar model are listed in 
Table~\ref{tab:abundances} (see also Bensby et al~\cite{bensby} for further
discussion).

Table~\ref{tab:solarvalues} lists the abundances we derive for the Sun from the
different lines using the methods that are described in more detail in
Sects.~\ref{sec:syre6300}, \ref{sec:syre6363}, and \ref{sec:syre7774} below.
For the \ion{O}{i} triplet we also give NLTE corrected abundances
(see Sect.~\ref{sec:syre7774}). There are quite large differences between the
abundances that the different indicators give. The [\ion{O}{i}] lines give
$\rm \epsilon(O)_{6300}$\,$=$\,8.71 and $\rm \epsilon(O)_{6363}$\,$=$\,9.06
while the \ion{O}{i} triplet lines give abundances slightly higher than
8.8\,dex if no NLTE correction is applied and slightly lower than 8.8\,dex
if corrected for NLTE effects according to the prescription of
Gratton et al.~(\cite{gratton2}). That the abundances from the \ion{O}{i}
triplet do not agree with the abundances from the [\ion{O}{i}] lines is not
unexpected due to the NLTE effects that the triplet lines are known to suffer
from (e.g. Kiselman~\cite{kiselman}).

%------------------------------------------------------------------------------
\begin{table}[t]
\caption{
        Our derived Solar oxygen abundances, given on the absolute abundance
        scale
        ($\rm \epsilon(O)$\,$=$\,$\log N_{\rm O}/N_{\rm H}$\,$+$\,12.0). All
        abundances have been determined under the assumption of LTE. NLTE
        corrections are done according to the prescription in
        Gratton et al.~(\cite{gratton}).
	In the third column we give the \ion{O}{i} triplet abundances corrected
	for NLTE.
	The two last columns give the differences
        between abundances from this work and the standard solar photospheric
        value, $\rm \epsilon(O)_{\sun}$\,$=$\,8.83,  from
        Grevesse \& Sauval~(\cite{grevesse}).
        }
\centering
\label{tab:solarvalues}
\begin{tabular}{ccccc}
\hline   \hline\noalign{\smallskip}
   Wavelength &  $\rm \epsilon(O)_{\sun}$ & $\rm \epsilon(O)_{\sun}$  & \multicolumn{2}{c}{Difference} \\
      (\AA )  &         LTE               &   NLTE-corr               &  LTE    &   NLTE               \\
\noalign{\smallskip}
\hline\noalign{\smallskip}
     6300.304 &   8.71         &       & $-0.12$   &           \\
     6363.776 &   9.06         &       & $+0.23$   &           \\
     7771.944 &   8.83         &  8.78 & $\pm0.00$ & $-0.05$   \\
     7774.166 &   8.88         &  8.83 & $+0.05$   & $\pm0.00$ \\
     7775.388 &   8.82         &  8.77 & $-0.01$   & $-0.06$   \\
\noalign{\smallskip}
\hline
\end{tabular}
\end{table}
%------------------------------------------------------------------------------

Lambert~(\cite{lambert}) noted that there is a possible weak CN blend in the 
[\ion{O}{i}]$_{6363}$ line and estimated its equivalent width to be 
$\sim$\,0.18\,m{\AA} in the Sun. 
Reducing our measured equivalent width by 
this amount results in an oxygen abundance that is 0.03\,dex lower, i.e. there 
is still almost a 0.3\,dex difference between abundances from the two forbidden 
lines. In order to bring down the abundance from the [\ion{O}{i}]$_{6363}$ to 
what we find for the [\ion{O}{i}]$_{6300}$ line we would have to reduce the 
equivalent width from 2.3\,m{\AA} to 1.0\,m{\AA}, i.e. the CN blend would have 
a strength of 1.3\,m{\AA} in the Sun (if this is the only blend). 
Another possible reason for the inconsistency between the 
abundances from the two forbidden lines is erroneous $\log gf$-values. 
However, the $\log gf$-values for these lines come from the same calculations 
(see Sect.~\ref{sec:atomdata}). If they are erroneous it is therefore most 
likely that both lines show deviations in the same direction and of equal size.
Further, the [\ion{O}{i}]$_{6363}$ line is located in the wing of a
\ion{Ca}{i} auto-ionization line (Mitchell \& Mohler~\cite{mitchell})
which might affect the strength, and thereby the derived oxygen abundance,
from the [\ion{O}{i}]$_{6363}$ line (see Sect.~\ref{sec:syre6363}). 

The standard solar photospheric oxygen abundance is 
$\rm \epsilon(O)_{\sun}$\,$=$\,8.83\,$\pm$\,0.06 
(Grevesse \& Sauval~\cite{grevesse}), However, depending on which 
spectral line that is used likely values range from $\sim$\,8.6\,dex to 
$\sim$\,8.9\,dex using one-dimensional (1-D) models (Asplund~\cite{asplund}). 
The new 3-D models give better agreement between the derived Solar abundance 
from different indicators (Asplund~\cite{asplund}). It should be noted that 
our solar oxygen abundance derived from the [\ion{O}{i}]$_{6300}$ line, using 
standard 1-D LTE models, agrees nicely with what 
Allende Prieto et al.~(\cite{allendeprieto}) found using 3-D models 
($\rm \epsilon(O)_{\sun}$\,$=$\,8.69).

Since we want to compare the oxygen abundance trends to those oxygen trends 
that can be found in other works we normalize our abundances individually, 
line by line, to the solar values that we derived. This is done by correcting 
the abundances for all stars by subtracting the amounts that are given in the 
two final columns in Table~\ref{tab:solarvalues}. This means that the Sun will 
have $\rm [O/H]$\,$=$\,0 independent of which line that is considered. It is 
these corrected oxygen abundances that we give in Table~\ref{tab:abundances}.

%------------------------------------------------------------------------------
\begin{figure}
\centering
\resizebox{\hsize}{!}{\includegraphics[bb=18 144 592 477,clip]{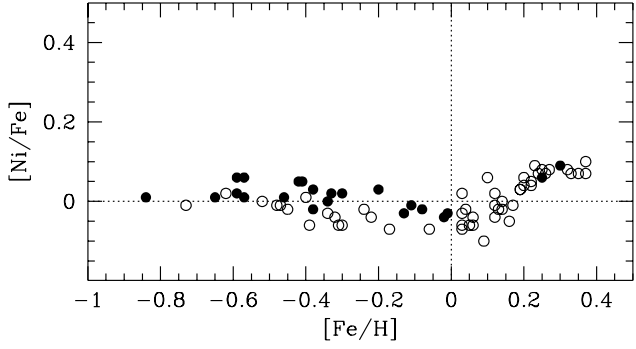}}
\caption{
        The observed Ni trend for our stellar sample (see
        Bensby et al.~\cite{bensby}). The six additional stars have been
        included as well (Bensby et al., in prep.). Thin and thick disk stars 
	are marked by open and filled circles, respectively.
         }
\label{fig:nitrend}
\end{figure}
%------------------------------------------------------------------------------

%==============================================================================
\subsection{Abundances from the {\rm [}\ion{O}{i}{\rm ]} line at 6300 {\AA}}  
	\label{sec:syre6300}

The oxygen abundance from the [\ion{O}{i}]$_{6300}$ line has been determined
for 63 of the 72 stars. In 60 cases we used CES spectra, in 2 cases FEROS
spectra, and in 1 case a UVES spectrum, as indicated in
Table~\ref{tab:abundances}. For 7 stars we were unable to use the
[\ion{O}{i}]$_{6300}$ line due to strong interference from the telluric
emission line (see Sect.~\ref{sec:reduction}). In two other cases where
abundances from this line are lacking we did not have CES spectra and the lower
quality of the FEROS spectra did not allow any abundance determination from
this line.

\begin{figure}
\centering
\resizebox{\hsize}{!}{\includegraphics[bb=18 144 592 718,clip]{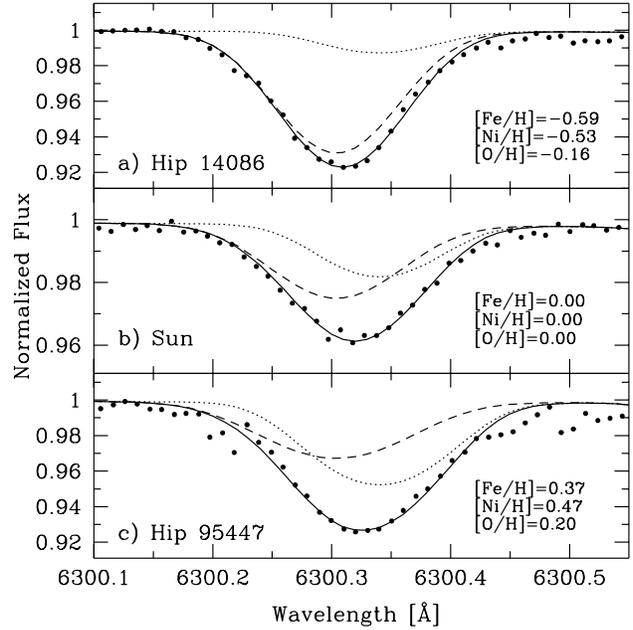}}
\caption{
        Synthesis of the [\ion{O}{i}]$_{6300}$ line for three stars. In each
        plot the observed spectrum is marked by solid circles and the synthetic
        spectra by lines: the [\ion{O}{i}] contribution by a dashed line; the
        contribution from the Ni lines by a dotted line; and the joint
        [\ion{O}{i}]\,-\,Ni contribution by a solid line. The solar spectrum
        is in this case a Moon spectrum. Abundances as indicated.
         }
\label{fig:lineprofile}
\end{figure}

The synthesis of the [\ion{O}{i}]$_{6300}$ line was done taking the blending
\ion{Ni}{i} lines into account. Figure~\ref{fig:nitrend} shows the observed Ni
trend for the stellar sample. It is obvious that [Ni/H] is not
varying in lockstep with [Fe/H] which is commonly assumed in the modeling of
the [\ion{O}{i}]$_{6300}$ line (e.g Nissen et al.~\cite{nissen2002}). At
[Fe/H]\,$\gtrsim$\,0 solar-scaled Ni abundances will underestimate the Ni
contribution to the line and consequently overestimate the final oxygen
abundance. However, at [Fe/H]\,$\lesssim$\,0 the [Ni/Fe]\,$\approx$\,0
assumption might be reasonably correct as the Ni contribution to the total line
strength decreases rapidly below [Fe/H]\,$=$\,0. As can be seen in
Fig.~\ref{fig:lineprofile}a the \ion{Ni}{i} contribution to the combined line
profile is small but not negligible for Hip 14086 which is a star that have
$\rm [Fe/H] = -0.59$. For the Sun the \ion{Ni}{i} contribution to the total
absorption is slightly less than 50\% (Fig.~\ref{fig:lineprofile}b) and for
Hip 95447 that has $\rm [Fe/H]=0.37$ the \ion{Ni}{i} contribution exceeds 50\%
of the combined [\ion{O}{i}]-\ion{Ni}{i} line profile
(Fig.~\ref{fig:lineprofile}c).

%------------------------------------------------------------------------------
%------------------------------------------------------------------------------

%------------------------------------------------------------------------------
\begin{figure}
 \resizebox{\hsize}{!}{\includegraphics[bb=18 144 592 718,clip]{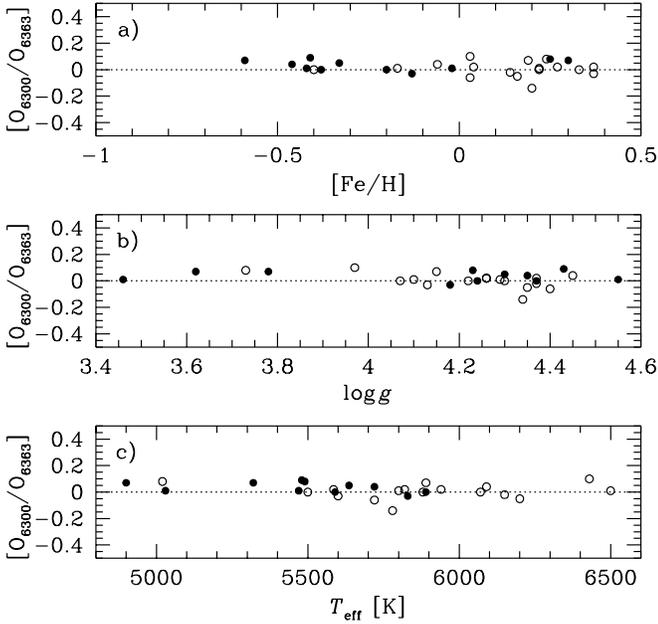}}
\caption{
        Difference between the oxygen abundances from the [\ion{O}{i}]$_{6300}$
        line and the [\ion{O}{i}]$_{6363}$ line versus [Fe/H], $\log g$, and
        $T_{\rm eff}$. Thin and thick disk stars are marked by open and filled
        circles, respectively.
         }
\label{fig:odiffs_6363}
\end{figure}
%------------------------------------------------------------------------------

To determine the oxygen abundance we first created a set of synthetic spectra
representative of different oxygen abundances (in steps of 0.03\,dex),
using the $v_{\rm rot-macro}$ determined in Sect.~\ref{sec:rotation}. Each
synthetic spectrum was compared to the observed spectrum and a corresponding
(un-normalized) $\chi^2$-value was calculated. In this way we got a
$\chi^2$-function versus oxygen abundance that was minimized to find the best
value of the oxygen abundance for each star. The corresponding abundance for
the \ion{Ni}{i} blends were kept at a fixed value and was gathered from
Bensby et al.~(\cite{bensby}). Tables~\ref{tab:abundances} and
\ref{tab:o7774abundances} lists the derived abundances.

%==============================================================================
\subsection{Abundances from the {\rm [}\ion{O}{i}{\rm ]} line at 6363 {\AA}}
        \label{sec:syre6363}

The [\ion{O}{i}]$_{6363}$ line is approximately 50\% weaker than the
[\ion{O}{i}]$_{6300}$ line and is located in the right wing of a wide
\ion{Ca}{i} auto-ionization line (Mitchell \& Mohler~\cite{mitchell}). 
This makes it difficult to synthesize this line well in our FEROS spectra. 
Instead we determined abundances from equivalent width
measurements. That the continuum is depressed is then of small concern since
the local continuum can be set at every measurement. 
However, depending on at which depth in the stellar atmosphere
this \ion{Ca}{i} auto-ionization line forms its effect on the equivalent 
width of the [\ion{O}{i}]$_{6363}$ line will vary. If it is formed 
below the [\ion{O}{i}]$_{6363}$ line
it will only contribute to the continuum flux, while if it forms higher up in 
the atmosphere than the [\ion{O}{i}]$_{6363}$ line
it will increase the strength of the [\ion{O}{i}]$_{6363}$ line.
If this latter is the case and if the estimation of
the strength of the CN blend by Lambert~(\cite{lambert}) is correct
(and also assuming that CN is the only blend), this
implies that the \ion{Ca}{i} auto-ionization line contributes
$\sim$\,1.1\,m{\AA} to the [\ion{O}{i}]$_{6363}$ line in the Sun.

Both [C/H] and [N/H] evolves in lockstep with [Fe/H] at high 
metallicities (Andersson \& Edvardsson~\cite{andersson};
Gustafsson et al.~\cite{gustafsson2}; Clegg et al.~\cite{clegg}; 
Shi et al.~\cite{shi}). From this it does not necessarily follow
that [CN/H] evolves like [Fe/H]. The amount of CN present in a stellar
atmosphere depends on its dissociation energy and the temperature in the
stellar atmosphere. The dissociation energy for CN is, however,
not well known (see discussion in de Laverny \& Gustafsson~\cite{laverny}).
But, as can be seen in Figs.~\ref{fig:odiffs_6363}a and c, there is no
trends in [O$_{6300}$/O$_{6363}$] with either [Fe/H] or $T_{\rm eff}$
which possibly indicates that [CN/H] evolves with [Fe/H] in our stars. 

Since we have neglected the CN blend (and other potential effects) 
in the abundance determination there is a
potential risk that the oxygen abundances from the [\ion{O}{i}]$_{6363}$ line
are {\it too high}. 
How it will affect the [O/Fe] vs [Fe/H] trends depends on how the CN abundance 
evolves with [Fe/H]. 
It seems that [CN/Fe]\,$\approx$\,0 for all [Fe/H] 
(if CN is the only blend) since the (normalized) 
oxygen abundances are the same for both of the forbidden lines 
(see Figs.~\ref{fig:odiffs_6363}a\,--\,c).
The effect on the derived [O/Fe] trends will therefore only be a systematic
vertical shift for all stars since we neglected the contribution from
the blend in our Solar analysis as well. 

Due to the lower
resolution and the lower signal-to-noise of the FEROS spectra compared to the
CES spectra we could only measure the equivalent width of the 
[\ion{O}{i}]$_{6363}$ line in 32
stars. A UVES spectrum was used in one case (see Table~\ref{tab:abundances}).
The measured equivalent widths for the [\ion{O}{i}]$_{6363}$ line are listed in
Table~\ref{tab:o7774abundances}.

%==============================================================================
\subsection{Abundances from the \ion{O}{i} triplet at 7774 {\AA}}
        \label{sec:syre7774}

Abundances were determined through equivalent width measurements of the
permitted \ion{O}{i} lines at 7771.95/7774.17/7775.39\,{\AA} in our FEROS 
spectra for 69
stars. The UVES spectra for the three additional stars could not be used since
the triplet lines are not present due to the setting of the spectrograph.

The triplet lines are known to give discrepant abundances under the
assumption of LTE compared to the [\ion{O}{i}]$_{6300}$ line 
(e.g. Eriksson \& Toft~\cite{eriksson};
Kiselman~\cite{kiselman}, \cite{kiselman2}; Gratton et al.~\cite{gratton2}). 
An LTE analysis will always give abundances that are overestimated. In our
analysis we correct the derived abundances for NLTE effects according to the
prescription of Gratton et al.~(\cite{gratton2}).
These non-LTE corrections are based on comparisons between abundances
obtained with statistical equilibrium calculations and abundances provided
by an LTE analysis with the Uppsala MARCS models.

%------------------------------------------------------------------------------
\begin{figure*}
\resizebox{\hsize}{!}{
        \includegraphics[bb=18 144 592 718,clip]{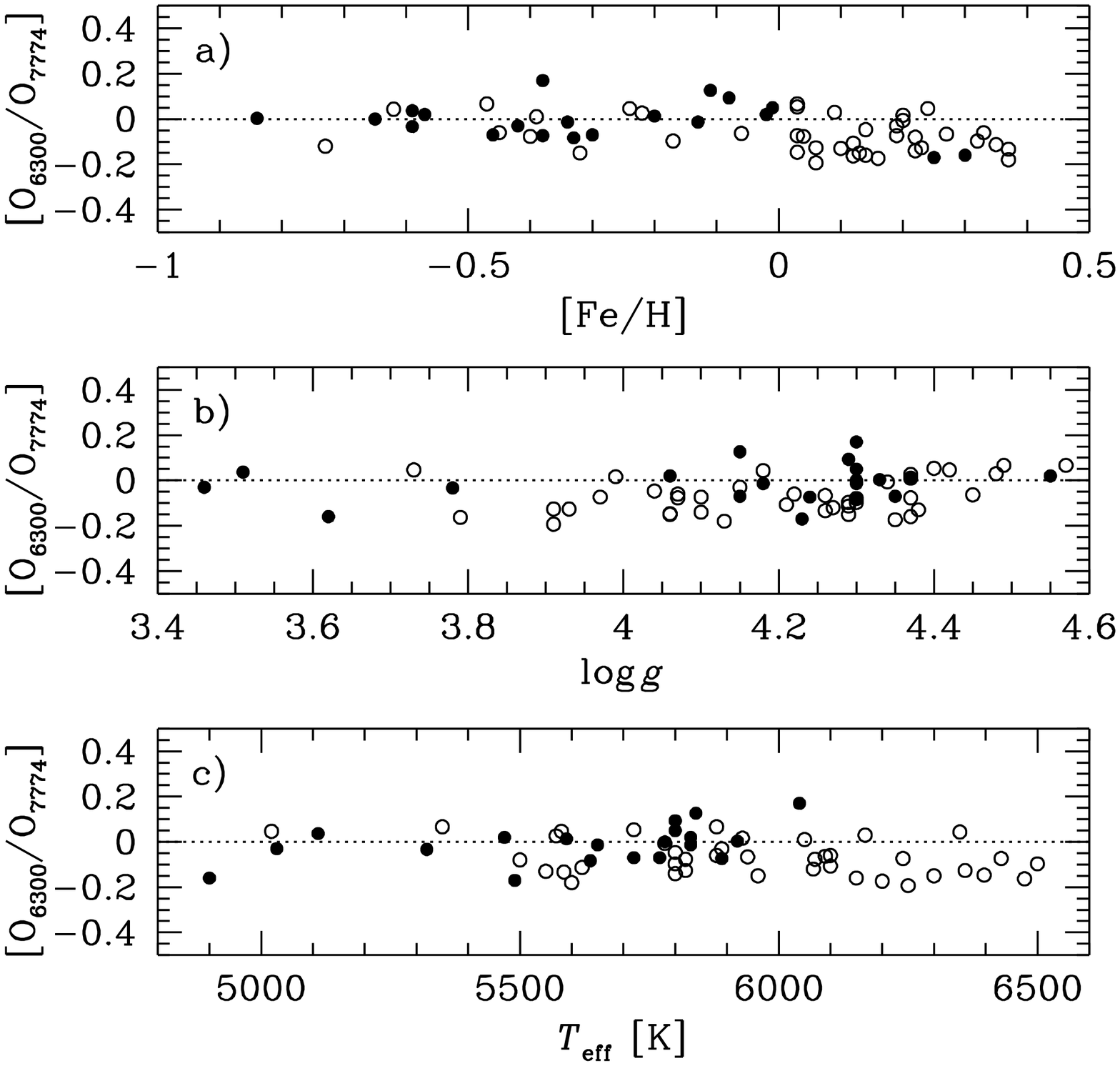} \hspace{10mm}
        \includegraphics[bb=18 144 592 718,clip]{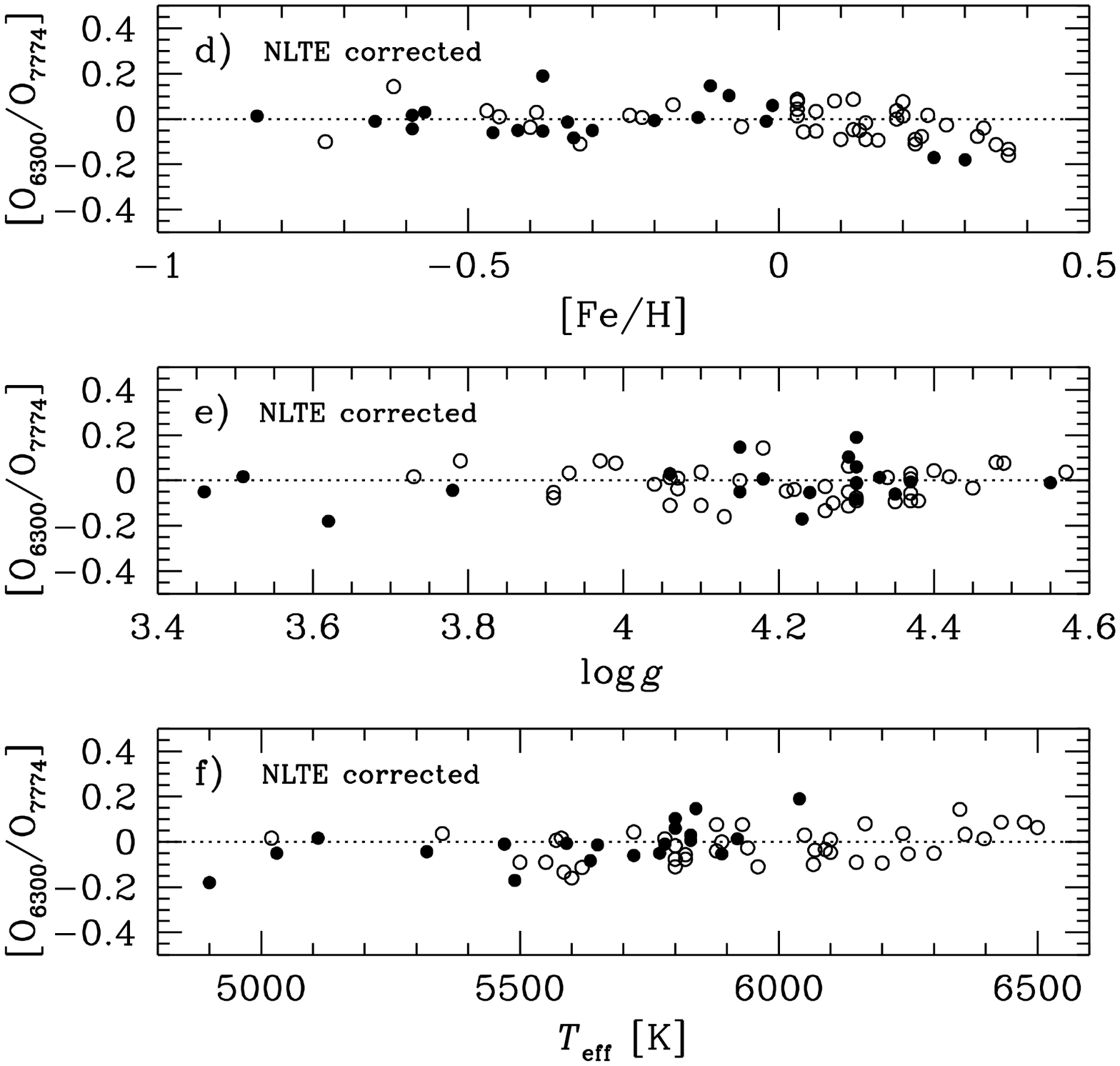}}
\caption{
        {\bf a)}--{\bf c)} Difference between the oxygen abundances from the
        [\ion{O}{i}]$_{6300}$ line and the \ion{O}{i} triplet at 7774\,{\AA}
        versus [Fe/H], $\log g$, and $T_{\rm eff}$. No NLTE corrections have
        been applied to the abundances from the triplet.
        {\bf d)}--{\bf f)} Difference between the oxygen abundances from the
        [\ion{O}{i}]$_{6300}$ line and the NLTE corrected abundances (according
        to the prescription in Gratton et al.~\cite{gratton2}) from the
        \ion{O}{i} triplet at 7774\,{\AA} versus [Fe/H], $\log g$, and
        $T_{\rm eff}$. Thin and thick disk stars are marked by open and filled
        circles, respectively.
         }
\label{fig:odiffs_7774}
\end{figure*}
%------------------------------------------------------------------------------

%------------------------------------------------------------------------------
\begin{table*}
\caption{
        Oxygen abundances for the 72 program stars. Column one gives the 
	Hipparcos number, columns 2\,--\,6 the derived oxygen abundances for 
	the different indicators relative to the photospheric oxygen abundance 
	($\rm \epsilon(O)_{\sun} = 8.83$) from 
	Grevesse \& Sauval~(\cite{grevesse}). Columns 7\,--\,11 give the 
	abundances relative to our solar abundances, 
	i.e. [O/H]$_{\sun}$\,$=$\,0 for all indicators. Columns 12\,--\,14 give 
	the abundances for the triplet that have been corrected for NLTE 
	effects according the the prescription in 
	Gratton et al.~(\cite{gratton}), and normalized to the new NLTE 
	corrected abundance for the Sun. Column 15 gives the NLTE correction 
	terms from Gratton et al.~(\cite{gratton}) (GCEG) that were applied to 
	the abundances in columns 12\,--\,14. Column 15 gives the empirical 
	NLTE correction we determined (Eqs.~\ref{eq:NLTEcorr} and 
	\ref{eq:NLTEcorr2}). Columns 16\,--\,19 give the measured 
	equivalent widths for the [\ion{O}{i}]$_{6363}$ and the triplet lines. 
	The full table is available in electronic form at the CDS via anonymous 
	ftp to \texttt{cdsarc.u-strasbg.fr} (130.79.128.5) or via 
	\texttt{cdsweb.u-strasbg.fr/cats/J.A+A.all.htx}
        }
\centering \scriptsize
\setlength{\tabcolsep}{1.6mm}
\label{tab:o7774abundances}
\begin{tabular}{rrrrrrrrrrrrrrrrrrrrr}
\hline   \hline\noalign{\smallskip}
Hip     &  \multicolumn{5}{c}{---------------------~[O/H]~---------------------}
        &  \multicolumn{5}{c}{------------------~[O/H]$_{\rm norm}$~------------------}
        &  \multicolumn{3}{c}{-------~[O/H]$^{\rm NLTE}_{\rm norm}$~-------}
        &  \multicolumn{2}{c}{NLTE-corr.}
        &  \multicolumn{4}{c}{----------~EW ~[m{\AA}]~----------}  \\
\noalign{\smallskip}
        &  6300
        &  6363
        &  7772
        &  7774
        &  7775
        &  6300
        &  6363
        &  7772
        &  7774
        &  7775
        &  7772
        &  7774
        &  7775
        &  GCEG
        &  TW
        &  6363
        &  7772
        &  7774
        &  7775   \\
\noalign{\smallskip}
\hline\noalign{\smallskip}
Sun     & $-0.12$ & $ 0.23$ & $ 0.00$ & $ 0.05$ & $-0.01$ & $ 0.00$ & $ 0.00$ & $ 0.00$ & $ 0.00$ & $ 0.00$ & $ 0.00$ & $ 0.00$ & $ 0.00$ & $-0.05$ & $-0.04$ &  2.3    &  68.3   &  61.9   &  46.4   \\
3142    & $-0.37$ &         & $-0.15$ & $-0.16$ & $-0.22$ & $-0.25$ &         & $-0.15$ & $-0.21$ & $-0.21$ & $-0.22$ & $-0.28$ & $-0.28$ & $-0.12$ & $-0.06$ &         &  88.3   &  77.1   &  58.6   \\
3170    &         &         & $-0.09$ & $-0.07$ & $ 0.02$ &         &         & $-0.09$ & $-0.12$ & $ 0.03$ & $-0.13$ & $-0.16$ & $-0.01$ & $-0.09$ & $-0.06$ &         &  83.2   &  74.5   &  65.8   \\
\vdots  & \vdots  & \vdots  & \vdots  & \vdots  & \vdots  & \vdots  & \vdots  & \vdots  & \vdots  & \vdots  & \vdots  & \vdots  & \vdots  & \vdots  & \vdots & \vdots  & \vdots  & \vdots  & \vdots  \\
\noalign{\smallskip}
\hline
\end{tabular}
\end{table*}
%------------------------------------------------------------------------------

Table~\ref{tab:o7774abundances} lists our measured equivalent widths as well as 
abundances for individual lines and NLTE corrections from 
Gratton et al.~(\cite{gratton2}). Table~\ref{tab:abundances} gives the mean 
abundances (both LTE and NLTE) from the three lines. 

In Figs.~\ref{fig:odiffs_7774}a\,--\,c we compare the abundances, not 
corrected for NLTE effects (but normalized to the Sun), from the triplet lines
to those from the [\ion{O}{i}]$_{6300}$ line for the 60 stars
that have both the [\ion{O}{i}]$_{6300}$ and the triplet lines analyzed
(compare Table~\ref{tab:abundances}). In summary, 
abundances from the triplet give higher abundances than what the forbidden 
line gives. The average difference and the standard deviation are 
$\rm \langle [O_{6300}/O_{7774}] \rangle$\,$=$\,$-0.049\pm0.084$. 
The largest differences are seen at the highest metallicities 
([Fe/H]\,$\gtrsim$\,0) and highest temperatures 
($T_{\rm eff}$\,$\gtrsim$\,6100\,K) and, possibly, for 
$\log g$\,$\gtrsim$\,4.4.
For the NLTE corrected abundances the average difference and the 
standard deviation is lower 
$\rm \langle [O_{6300}/O_{7774}] \rangle$\,$=$\,$-0.015$\,$\pm$\,0.077
(see Figs.~\ref{fig:odiffs_7774}d\,--\,f). The trends with [Fe/H] and 
$\log g$ are 
still present but somewhat less prominent and the trend with $T_{\rm eff}$ has 
changed significantly. The triplet lines now give increasingly lower abundances 
when going to higher $T_{\rm eff}$. The NLTE corrections of 
Gratton et al.~(\cite{gratton2}) therefore could 
be somewhat overestimated for the parameter space that our stars span.

A weighted linear regression\footnote{For a description of the algorithm see 
the manual entry for the``robustfit'' function in the {\sc Matlab} 
{\it Statistics Toolbox User's Guide} ({\tt www.mathworks.com})}
with $\log T_{\rm eff}$, $\log g$, and $[\mathrm{Fe}/\mathrm{H}]$ as the 
dependent variables and the difference between the oxygen abundances from the 
[\ion{O}{i}]$_{6300}$ line and the triplet lines ([O$_{6300}$/O$_{7774}$]) 
as the independent gives:
%------------------------------------------------------------------------------
\begin{eqnarray}
 \Delta_{\rm NLTE}^{7774} \equiv [\mathrm{O}_{6300}/\mathrm{O}_{7774}]   =  
	& - & 0.46\,(\pm0.02) \nonumber\\ 
        & - & 0.89\,(\pm0.37) \cdot\log (T_{\rm eff}/T_{\rm eff, \sun}) \nonumber\\
        & + & 0.10\,(\pm0.04) \cdot\log (g/g_{\sun}) \nonumber\\
        & - & 0.10\,(\pm0.03) \cdot [\mathrm{Fe}/\mathrm{H}].
\label{eq:NLTEcorr}
\end{eqnarray}
%------------------------------------------------------------------------------
This empirical correction can
then be applied to the abundances from the \ion{O}{i} triplet:
%------------------------------------------------------------------------------
\begin{equation}
[\rm{O}/\rm{H}]_{\rm NLTE}^{7774} = [\rm{O}/\rm{H}]_{\rm LTE}^{7774} + 
                                    \Delta_{\rm NLTE}^{7774}.
\label{eq:NLTEcorr2}
\end{equation}
%------------------------------------------------------------------------------

The derived relationship is valid for stars with: 
5000\,$\lesssim$\,$T_{\rm eff}$\,$\lesssim$\,6500, 
3.5\,$\lesssim$\,$\log g$\,$\lesssim$\,4.5,
and $-1$\,$\lesssim$\,${\rm [Fe/H]}$\,$\lesssim$\,0.5.

%==============================================================================
\section{Errors in resulting abundances} \label{sec:errors}

%------------------------------------------------------------------------------
\begin{table}
 \centering
 \caption{
        Estimates of the effects on the derived abundances due to internal 
	(random) errors for three stars. When calculating 
	$\Delta W_{\lambda}/\sqrt{N}$ for the Fe abundances we have assumed
        $\Delta W_{\lambda}$\,$=$\,5\,\% for Hip 88622 and Hip 82588 and 
	$\Delta W_{\lambda}$\,$=$\,10\,\% for Hip 103682 
	(see Bensby et al.~\cite{bensby}). For the [\ion{O}{i}]$_{6363}$ 
	abundances we adopted $\Delta W_{\lambda}$\,$=$\,10\,\% and for the 
	\ion{O}{i} triplet abundances $\Delta W_{\lambda}$\,$=$\,5\,\% for all 
	three stars. The total random uncertainties ($\sigma_{\rm rand}$) were 
	calculated assuming the individual errors to be uncorrelated. The final 
	line gives the average of the total random uncertainties for the three 
	stars.
            }
\label{tab:errors}
\centering \scriptsize
\setlength{\tabcolsep}{1.5mm}
\begin{tabular}{lrrrrrrr}
\hline \hline\noalign{\smallskip}
      &
      & \multicolumn{2}{c}{-----~6300~-----}
      & \multicolumn{2}{c}{-----~6363~-----}
      & \multicolumn{2}{c}{-----~7774~-----} \\
\noalign{\smallskip}
      & $\rm \left[\frac{\ion{Fe}{i}}{H}\right]$
      & $\rm \left[\frac{\ion{O}{i}}{H}\right]$
      & $\rm \left[\frac{\ion{O}{i}}{Fe}\right]$
      & $\rm \left[\frac{\ion{O}{i}}{H}\right]$
      & $\rm \left[\frac{\ion{O}{i}}{Fe}\right]$
      & $\rm \left[\frac{\ion{O}{i}}{H}\right]$
      & $\rm \left[\frac{\ion{O}{i}}{Fe}\right]$ \\
 \noalign{\smallskip}
 \hline\noalign{\smallskip}
           & \multicolumn{7}{c}{\bf Hip 88622} \\
 \noalign{\smallskip}
 $\Delta T_{\rm eff} = +70$ K                 & $ 0.06$ & $ 0.02$ & $-0.04$ & $ 0.02$ & $-0.04$ & $-0.06$ & $-0.12$ \\
 $\Delta \log g = +0.1$                       & $-0.01$ & $ 0.04$ & $ 0.05$ & $ 0.05$ & $ 0.06$ & $ 0.02$ & $ 0.01$ \\
 $\Delta \xi_{\rm t} = +0.15$ km~s$^{-1}$     & $-0.02$ & $ 0.00$ & $ 0.02$ & $ 0.00$ & $ 0.02$ & $-0.01$ & $ 0.01$ \\
 $\rm \Delta [Fe/H] = +0.1$                   & $ 0.01$ & $ 0.03$ & $ 0.02$ & $ 0.03$ & $ 0.02$ & $ 0.01$ & $ 0.00$ \\
 $\rm \Delta \delta\gamma_{6} = +50\%$        & $ 0.00$ & $ 0.00$ & $ 0.00$ & $ 0.00$ & $ 0.00$ & $ 0.00$ & $ 0.00$ \\
 $\Delta W_{\lambda}/\sqrt{N}$                & $ 0.00$ &         &         & $ 0.04$ & $ 0.04$ & $ 0.01$ & $ 0.01$ \\
 $\rm \sigma_{\rm form}(Ni) = 0.09/\sqrt{49}$ &         & $ 0.00$ & $ 0.00$ &         &         &         &         \\
 \noalign{\smallskip}
 $\sigma_{\rm rand}$                          & $ 0.06$ & $ 0.05$ & $ 0.07$ & $ 0.07$ & $ 0.09$ & $ 0.07$ & $ 0.12$ \\
 \noalign{\smallskip}
 \hline

 \noalign{\smallskip}
           & \multicolumn{7}{c}{\bf Hip 82588} \\
\noalign{\smallskip}
 $\Delta T_{\rm eff} = +70$ K                 & $ 0.05$ & $ 0.01$ & $-0.04$ & $ 0.01$ & $-0.04$ & $-0.08$ & $-0.13$ \\
 $\Delta \log g = +0.1$                       & $-0.01$ & $ 0.04$ & $ 0.05$ & $ 0.05$ & $ 0.06$ & $ 0.02$ & $ 0.03$ \\
 $\Delta \xi_{\rm t} = +0.15$ km~s$^{-1}$     & $-0.03$ & $ 0.00$ & $ 0.03$ & $ 0.00$ & $ 0.03$ & $-0.01$ & $ 0.02$ \\
 $\rm \Delta [Fe/H] = +0.1$                   & $ 0.01$ & $ 0.04$ & $ 0.03$ & $ 0.03$ & $ 0.02$ & $ 0.02$ & $ 0.01$ \\
 $\rm \Delta \delta\gamma_{6} = +50\%$        & $-0.01$ & $ 0.00$ & $ 0.01$ & $ 0.00$ & $ 0.01$ & $ 0.00$ & $ 0.01$ \\
 $\Delta W_{\lambda}/\sqrt{N}$                & $ 0.00$ &         &         & $ 0.04$ & $ 0.04$ & $ 0.01$ & $ 0.01$ \\
 $\rm \sigma_{\rm form}(Ni) = 0.08/\sqrt{54}$ &         & $-0.01$ & $ 0.01$ &         &         &         &         \\
 \noalign{\smallskip}
 $\sigma_{\rm rand}$                          & $ 0.06$ & $ 0.06$ & $ 0.08$ & $ 0.07$ & $ 0.08$ & $ 0.09$ & $ 0.13$ \\
 \noalign{\smallskip}
 \hline

 \noalign{\smallskip}
           & \multicolumn{7}{c}{\bf Hip 103682} \\
 \noalign{\smallskip}
 $\Delta T_{\rm eff} = +70$ K                 & $ 0.05$ & $ 0.02$ & $-0.03$ & $ 0.00$ & $-0.05$ & $-0.06$ & $-0.11$ \\
 $\Delta \log g = +0.1$                       & $-0.01$ & $ 0.05$ & $ 0.06$ & $ 0.04$ & $ 0.05$ & $ 0.02$ & $ 0.03$ \\
 $\Delta \xi_{\rm t} = +0.15$ km~s$^{-1}$     & $-0.05$ & $ 0.00$ & $ 0.05$ & $ 0.00$ & $ 0.05$ & $-0.01$ & $ 0.04$ \\
 $\rm \Delta [Fe/H] = +0.1$                   & $ 0.00$ & $ 0.03$ & $ 0.03$ & $ 0.03$ & $ 0.03$ & $ 0.01$ & $ 0.01$ \\
 $\rm \Delta \delta\gamma_{6} = +50\%$        & $ 0.00$ & $ 0.00$ & $ 0.00$ & $ 0.00$ & $ 0.00$ & $ 0.00$ & $ 0.00$ \\
 $\Delta W_{\lambda}/\sqrt{N}$                & $ 0.00$ &         &         & $ 0.04$ & $ 0.04$ & $ 0.01$ & $ 0.01$ \\
 $\rm \sigma_{\rm form}(Ni) = 0.11/\sqrt{54}$ &         & $-0.01$ & $ 0.01$ &         &         &         &         \\
 \noalign{\smallskip}
 $\sigma_{\rm rand}$                          & $ 0.07$ & $ 0.06$ & $ 0.09$ & $ 0.06$ & $ 0.10$ & $ 0.07$ & $ 0.12$ \\
 \noalign{\smallskip}
 \hline
 \noalign{\smallskip}
 $\langle \sigma_{\rm rand}\rangle$           & $ 0.06$ & $ 0.06$ & $ 0.08$ & $ 0.07$ & $ 0.09$ & $ 0.08$ & $ 0.12$ \\
 \noalign{\smallskip}
 \hline
 \end{tabular}
\end{table}
%------------------------------------------------------------------------------

The random (internal) errors in the derived oxygen abundances can have several 
causes. Effective temperature, surface gravity, microturbulence, measured 
equivalent widths, and atomic data are the main contributors. 
In estimating the total error we assume that the errors arising from
each of the sources are independent. The error due to each parameter is 
estimated by varying the parameter in question.

As in Bensby et al.~(\cite{bensby}) we estimate the effects of the 
uncertainties in these parameters on the derived oxygen abundances by 
increasing $T_{\rm eff}$ by 70\,K, $\log g$ by 0.1\,dex, microturbulence 
($\xi_{\rm t}$) by 0.15\,km\,s$^{-1}$, [Fe/H] by 0.1\,dex, correction factor 
($\delta\gamma_6$) to the classic Uns\"old by 50\%, and the measured equivalent 
widths by 5\,--\,10\,\%. This uncertainty in the measured equivalent widths 
correspond to an uncertainty in the derived abundances of 0.02\,--\,0.04\,dex. 

If the abundance is derived from $N$ spectral lines the uncertainty should be 
reduced by a factor ${N}^{1/2}$ to give the formal error in the mean. For the 
[\ion{O}{i}]$_{6363}$ line we adopt the higher value (0.04\,dex) since the line 
is very weak and difficult to measure. For the \ion{O}{i} triplet abundance we 
adopt the lower value ($0.02/\sqrt{3}$\,dex) since the three lines are strong 
and easy to measure.

The uncertainty in the abundance from the [\ion{O}{i}]$_{6300}$ line is also 
dependent on the accuracy of the derived Ni abundance. From the discussion in 
Sect.~\ref{sec:syre6300} it is also clear that stars at high [Fe/H] are more 
affected by this than those at low [Fe/H]. To estimate these uncertainties we 
vary the Ni abundance according to the observed formal error in the mean of the 
Ni abundance (see Bensby et al.~\cite{bensby}) and then derive new oxygen 
abundances from line synthesis.

These error estimates are exemplified for three stars and are listed in 
Table~\ref{tab:errors}. It should be noted that the error estimates for the 
damping constants only apply to those lines marked by a ``U" in 
Table~\ref{tab:linelist}. The total internal (random) errors are typically 
0.06\,dex in [O/H]$_{6300}$ and [O/H]$_{6363}$, and 0.08\,dex in 
[O/H]$_{7774}$.

\begin{figure*}
 \resizebox{\hsize}{!}{
	\includegraphics[bb=18 144 592 718,clip]{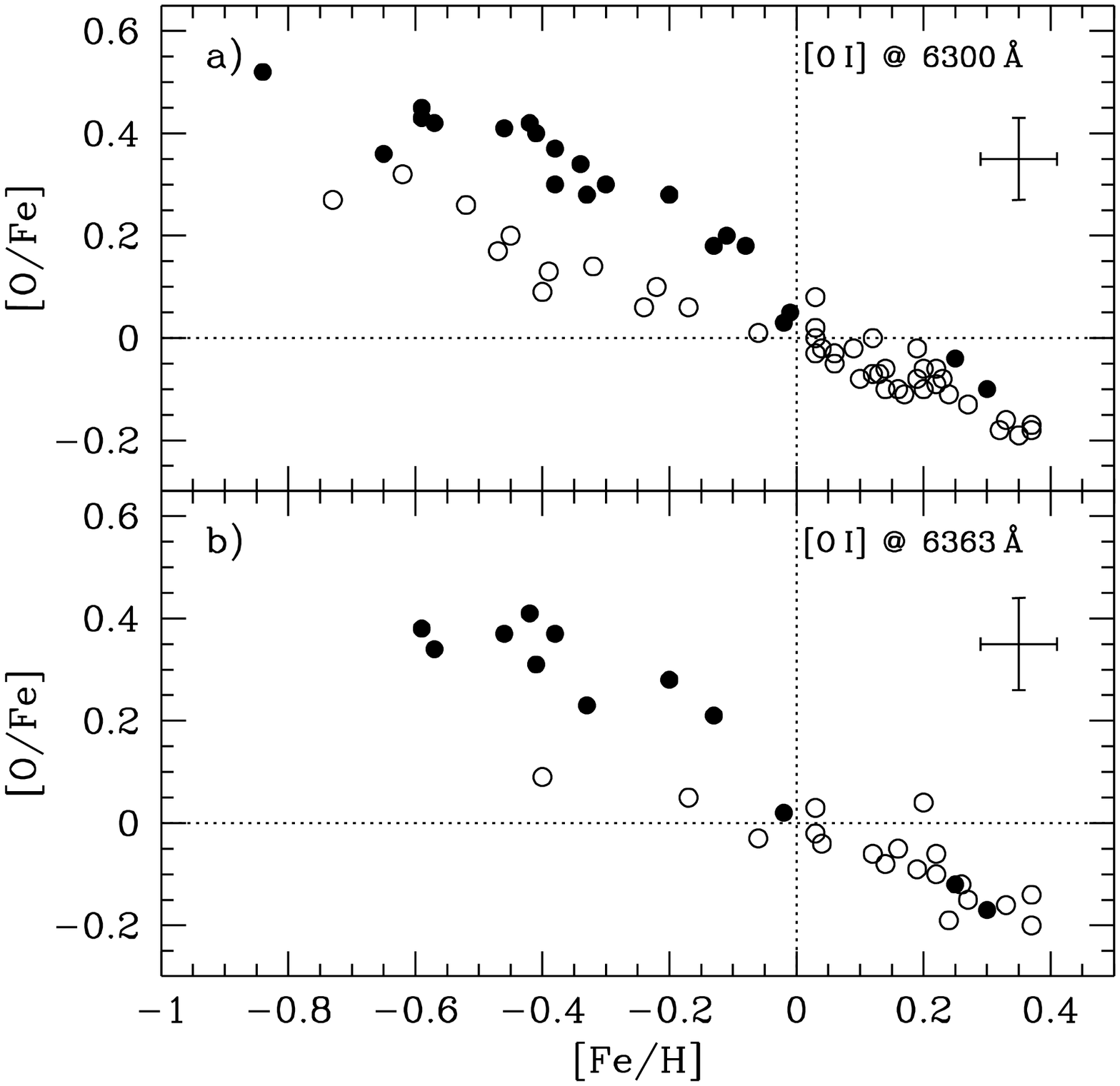} \hspace{10mm}
	\includegraphics[bb=18 144 592 718,clip]{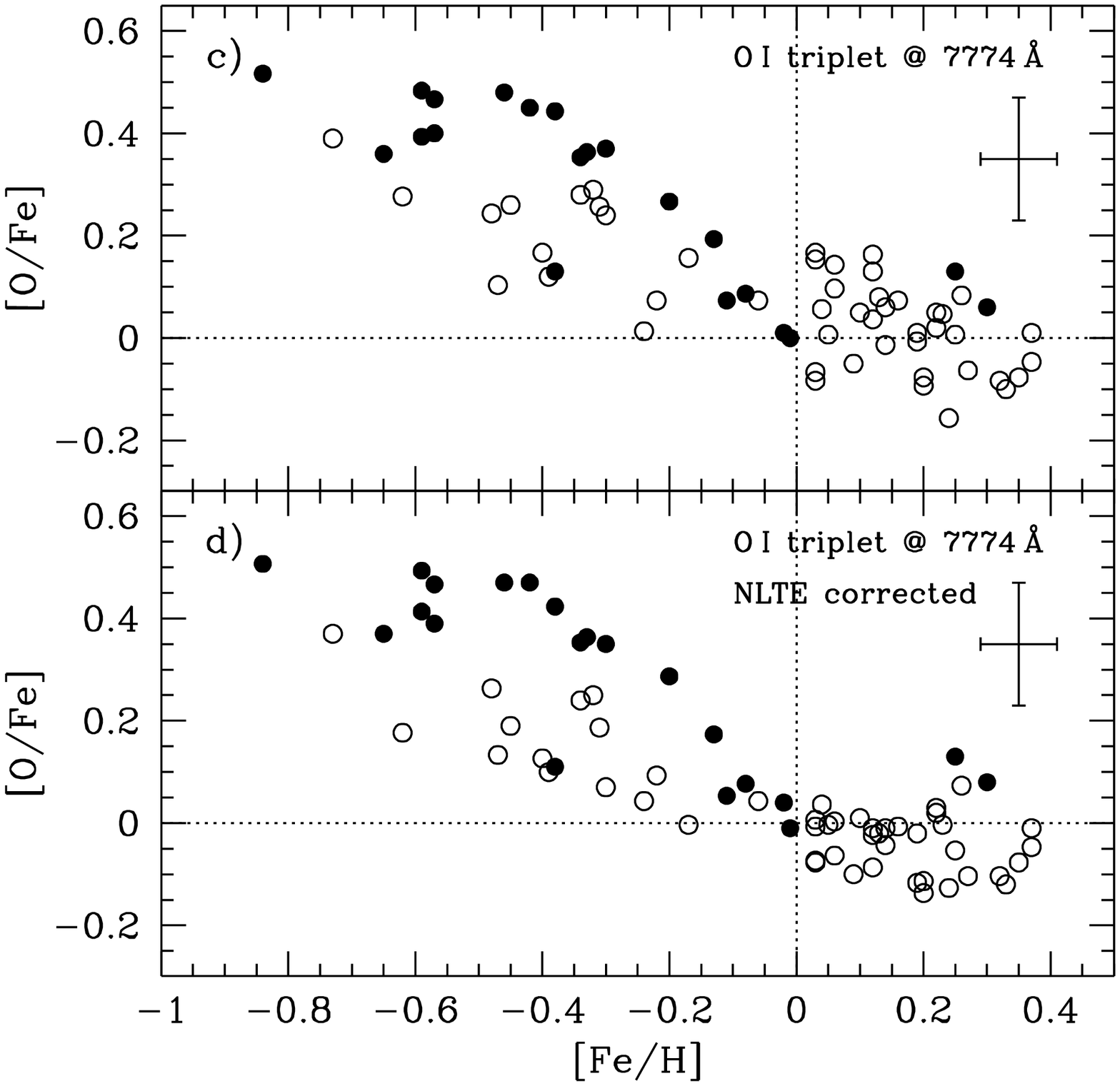}}
\caption{
        The oxygen trends for our stellar sample as it appears when using; 
	{\bf a)} the [\ion{O}{i}]$_{6300}$ line, {\bf b)} the 
	[\ion{O}{i}]$_{6363}$ line, {\bf c)} the \ion{O}{i} triplet at 
	7774\,{\AA}, and {\bf d)} the \ion{O}{i} triplet at 7774\,{\AA} 
	corrected for NLTE effects according to the prescription in 
	Gratton et al.~(\cite{gratton2}) (but see discussion in 
	Sect.~\ref{sec:syre7774}). 
	Thin and thick disk stars are marked 
	by open and filled circles, respectively. The error-bars in the upper 
	right corner are the total random errors (see Sect.~\ref{sec:errors} 
	and Table~\ref{tab:errors}).
         }
\label{fig:o_trends}
\end{figure*}

We note that the oxygen trends in Fig.~\ref{fig:o_trends}a and b are very
tight and that the two populations separate very well. That this separation
should be so distinct and the {\it internal} errors as large as the ones
estimated in this section seems unlikely. We therefore conclude that our error 
estimates in Table~\ref{tab:errors} are upper limits to the internal errors.

%------------------------------------------------------------------------------
%------------------------------------------------------------------------------

%==============================================================================
\section{[O/Fe] vs [Fe/H] trends}   \label{sec:abund_trends}

Our oxygen trends derived from the [\ion{O}{i}]$_{6300}$ line are shown in 
Fig.~\ref{fig:o_trends}a. The thin and thick disk trends are well separated and 
distinct. At [Fe/H]\,$<$\,$-0.4$ the [O/Fe] trend is nearly flat for the thick 
disk stars, showing an oxygen overabundance of [O/Fe]\,$\approx$\,0.4. Then, at 
[Fe/H]\,$\approx$\,$-$0.4, there is a downward trend towards solar values. This 
trend is in good agreement with what we found for other $\alpha$-elements 
(Mg, Si, Ca, and Ti) (see Bensby et al.~\cite{bensby}; 
Feltzing et al.~\cite{feltzing}). As we already noted then, this turn-over or 
``knee" is most likely an indication of contribution from SN\,Ia to the 
chemical enrichment of the interstellar medium.

The thin disk [O/Fe] trend is very well defined with a small internal scatter. 
It shows an overabundance of oxygen of [O/Fe]\,$\approx$\,0.3 at 
[Fe/H]\,$\approx$\,$-0.7$ and continues linearly down to an underabundance of 
[O/Fe]\,$\approx$\,$-0.2$ at [Fe/H]\,$\approx$\,0.4. Such a trend has 
previously only been seen by Castro et al.~(\cite{castro}) who analyzed the 
[\ion{O}{i}]$_{6300}$ line in seven stars with 
+0.1\,$\lesssim$\,[Fe/H]\,$\lesssim$\,+0.5, and by 
Chen et al.~(\cite{chen2003}) who analyzed the \ion{O}{i} triplet in 15 stars 
with $-0.1$\,$\lesssim$\,[Fe/H]\,$\lesssim$\,+0.5. However, only seven stars in 
the Chen et al.~(\cite{chen2003}) study had [Fe/H]\,$>$\,+0.1.

Compared to the abundance trends obtained from the [\ion{O}{i}]$_{6300}$ line, 
Fig.~\ref{fig:o_trends}a, the trends based on the [\ion{O}{i}]$_{6363}$ line is 
practically identical even though the number of stars is considerably smaller. 
Since the [\ion{O}{i}]$_{6363}$ line of a star is not affected by its 
Ni abundance 
this provides an independent test of our analysis of the 
[\ion{O}{i}]$_{6300}$ line. We therefore believe that the downward trend in 
[O/Fe] that we see at [Fe/H]\,$>$\,0 as well as the separation of the thin and 
thick disk trends at [Fe/H]\,$<$\,0 are real. 

In Fig.~\ref{fig:o_trends}c we show the abundance trends derived from the 
\ion{O}{i} triplet that have not been corrected for NLTE effects. At 
[Fe/H]\,$<$\,0 the trends for the thin and the thick disks are essentially the 
same as we saw from the forbidden lines (Figs.~\ref{fig:o_trends}a 
and b) suggesting that the NLTE effects might not be so severe at these 
metallicities (see also Fig.~\ref{fig:odiffs_7774}a). However, the scatter is 
significantly larger, especially for the thin disk. At super-solar [Fe/H] there 
is a large scatter and no clear sign of the downward trend that we see in 
the [\ion{O}{i}] lines. The [O/Fe] trend based on the NLTE 
corrected (according to the prescription by Gratton et al.~\cite{gratton2}) 
\ion{O}{i} abundances is shown in Fig.~\ref{fig:o_trends}d. While not much 
have changed at [Fe/H]\,$<$\,0, except for a somewhat more well-defined thin 
disk trend, the scatter is considerably reduced at [Fe/H]\,$>$\,0 and with a 
marginal downward trend in [O/Fe]. However, the forbidden lines always give 
superiorly defined trends.

%==============================================================================
\section{Discussion}  \label{sec:implications}

In Fig.~\ref{fig:syre} we show the oxygen trends, both the common
[O/Fe] vs. [Fe/H] trend and the nowadays often used [Fe/H] vs. [O/H] trend,
with all 72 stars in the sample included. The abundances from the
[\ion{O}{i}]$_{6300}$ line have been used whenever available. If not, our
first choice are the abundances from the [\ion{O}{i}]$_{6363}$ line and second
the NLTE corrected (according to the prescription of
Gratton et al.~\cite{gratton2}) abundances from the \ion{O}{i} triplet.

%==============================================================================
\subsection{Interpretation of the oxygen trends in the thin and thick disks}

%==============================================================================
\subsubsection{Strengthening of the trends} \label{sec:strengthening}

\begin{figure}
 \resizebox{\hsize}{!}{ 
        \includegraphics[bb=18 144 592 475,clip]{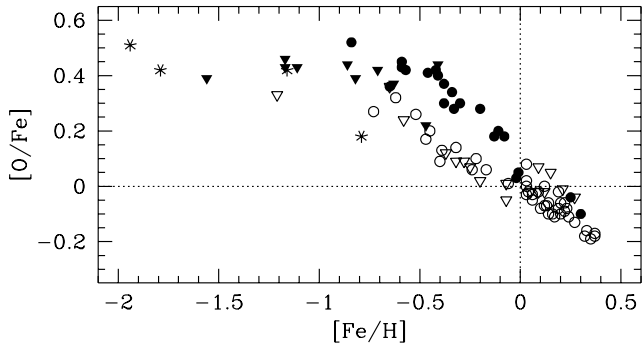}}
\caption{
        Oxygen trends based on the [\ion{O}{i}]$_{6300}$ line
        for our stars and from the study by Nissen et al.~(\cite{nissen2002}).
        The Nissen et al.~(\cite{nissen2002}) thin and thick
        disk stars are marked by open and filled triangles, respectively,
        and their halo stars by asterisks.
        Our thin and thick disk
        stars are marked by open and filled circles, respectively.
         }
\label{fig:syre_nissen}
\end{figure}

Nissen et al.~(\cite{nissen2002}) analyzed oxygen in 41 stars. Out of those,
31 stars have kinematic data available. For these we have calculated their 
TD/D and TD/H ratios and selected those with TD/D\,$>$\,1 and TD/H\,$>$\,1 as 
thick disk stars and those with TD/D\,$<$\,1 and TD/H\,$>$\,1 as thin disk 
stars. Four stars that have TD/D\,$>$\,1 and TD/H\,$<$\,1 were
consequently interpreted as halo stars. Apart from the 4 halo stars this 
left us with 10 thick disk and 17 thin disk stars from
Nissen et al.~(\cite{nissen2002}).
In Fig.~\ref{fig:syre_nissen} we add these stars to our own data set.

\begin{figure*}
\centering
\resizebox{\hsize}{!}{\includegraphics[angle=-90]{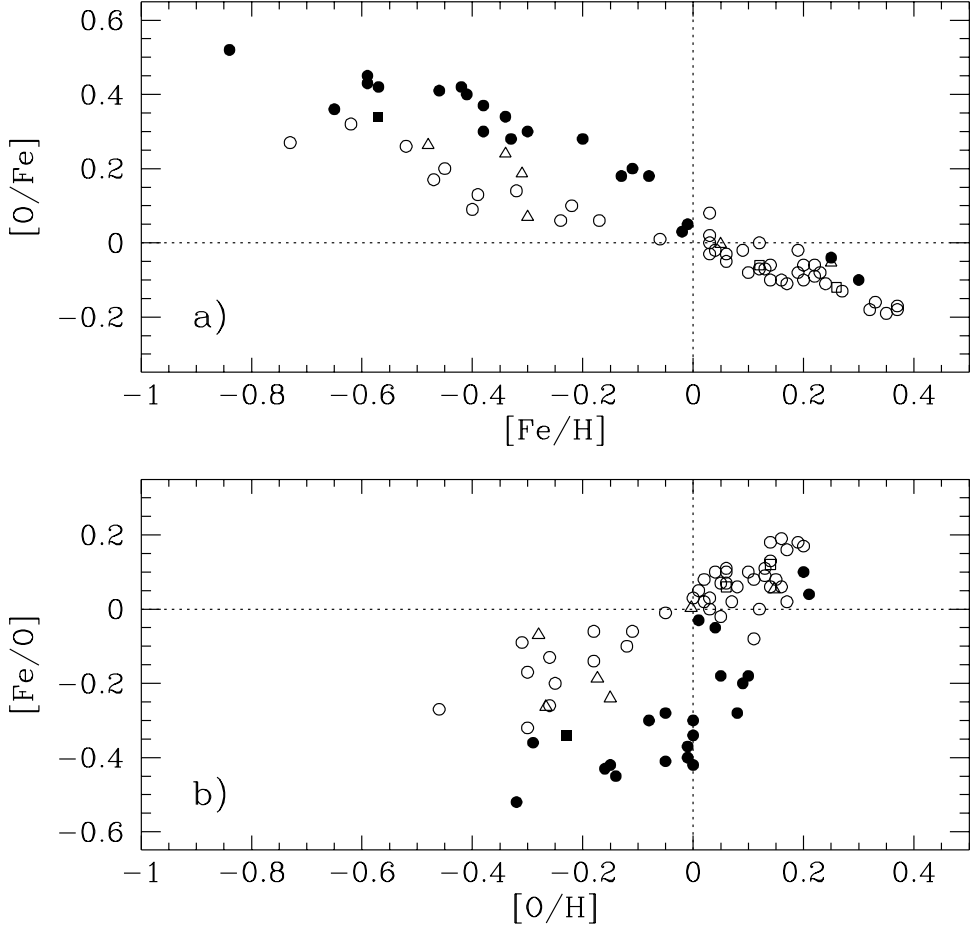}}
\caption{
        The oxygen trend, as it appears when all 72 stars are included.
        {\bf a)} shows the [O/Fe] vs. [Fe/H] trend and {\bf b)}
        the [Fe/O] vs. [O/H] trend. Abundances derived from the
        [\ion{O}{i}]$_{6300}$ line are marked by circles, abundances from the
        [\ion{O}{i}]$_{6363}$ line by squares, and NLTE corrected (according
        to the prescription in Gratton et al.~\cite{gratton}) abundances from
        the \ion{O}{i} triplet at 7774\,{\AA} by triangles. Thin and thick disk
        stars are marked by open and filled symbols, respectively.
         }
\label{fig:syre}
\end{figure*}

The thick and thin disk trends seen in Fig.~\ref{fig:syre}a are
further strengthened by the inclusion of the stars from
Nissen et al.~(\cite{nissen2002}) (see Fig.~\ref{fig:syre_nissen}). 
Especially so the flat trend in the
thick disk up till [Fe/H]\,$=$\,$-0.5$ and the break in [O/Fe] at
[Fe/H]\,$\approx$\,$-0.4$. The thin disk
trend at [Fe/H]\,$\lesssim$\,$0$ is nicely supported too. 
Intriguingly, the most metal-rich halo star
is low in [O/Fe] compared to thick disk stars at the same metallicity,
which was also found by Nissen \& Schuster~(\cite{schuster}). They speculated
that these anomalous halo stars can have been accreted from dwarf galaxies
that have a different chemical histories than the Galactic disks.

For the metal-rich disk ([Fe/H]\,$>$\,0) the Nissen et al.~(\cite{nissen2002})
stars have higher oxygen abundances than ours and level out to a constant
value of [O/Fe]\,$\sim$\,0.
The Nissen et al.~(\cite{nissen2002}) stars include a re-analysis of the
Nissen \& Edvardsson~(\cite{nissen1992}) data that did not account for
the Ni blend in the [\ion{O}{i}]$_{6300}$ line. 
Nissen et al.~(\cite{nissen2002})
calculated the equivalent width of the Ni blend assuming solar-scaled Ni
abundances (i.e. [Ni/Fe]\,$=$\,0) for all metallicities (which is not
consistent with what we see in our Ni trends, Fig.~\ref{fig:nitrend},
but is consistent to earlier studies e.g. Edvardsson et al.~\cite{edvardsson}). 
Then they
subtracted this calculated Ni contribution from the observed equivalent width 
of the [\ion{O}{i}] line and then re-calculated the oxygen abundances. Apart 
from an absolute shift in the oxygen abundances, their new results showed 
essentially no differences compared to the abundance trends in 
Nissen \& Edvardsson~(\cite{nissen1992}). This is not surprising since the 
assumption that [Ni/Fe]\,$=$\,0 for all [Fe/H] mainly results in a systematic 
shift of all abundances by a certain amount and will not affect the behaviour 
of the [O/Fe] trend. Hence the Nissen et al.~(\cite{nissen2002}) 
[O/Fe] trend still level out at [Fe/H]\,$>$\,0.

%------------------------------------------------------------------------------
%------------------------------------------------------------------------------

%==============================================================================
\subsubsection{Evolution of the thick disk}

In Bensby et al.~(\cite{bensby}) and Feltzing et al.~(\cite{feltzing}) we
discuss different formation scenarios for the thick disk. Based on our
elemental abundance trends in the thin and thick disks for the other
$\alpha$-elements (Mg, Si, Ca, and Ti) we concluded that it is very likely that
the thin and thick disks formed at different epochs. This conclusion is further
supported by the different mean stellar ages that we found from isochrones
for the thin and
thick disk stars. The average ages are (including all stars in this study)
4.9\,$\pm$\,2.8 and 10.8\,$\pm$\,4.3 for our thin and thick disk stellar
samples, respectively. However, our results could be accommodated both in a
dissipational collapse scenario (either fast or slow) or in a
merger/interacting scenario for the formation of the thick disk. 
However, evidence from extra-galactic studies of edge-on spiral galaxies
(e.g.\ Reshetnikov \& Combes~\cite{reshetnikov};
Schwarzkopf \& Dettmar~\cite{schwarzkopf};
Dalcanton \& Bernstein~\cite{dalcanton})
appear to indicate that thick disks are more common in galaxies that are
or have experienced mergers or interaction. Thus, at the moment the most
plausible formation scenario for the thick disk is a merger event 
(e.g. Quinn et al.~\cite{quinn}) or an interaction with a companion galaxy 
(e.g. Kroupa~\cite{kroupa})

The thick disk shows the signatures of enrichment from SN\,Ia to the 
interstellar medium. This is clearly seen in plots where [O/Fe] is plotted
versus [Fe/H] (Fig.~\ref{fig:syre}a and \ref{fig:syre_nissen}). The ``knee"
in [O/Fe] at [Fe/H]\,$\approx$\,$-0.4$ indicates when the SN\,Ia rate peaks
and thereby also the peak in the enrichment of Fe from these events. 
That the ``knee" is present at all also indicates that the
star formation in the early thick disk has been vigorous.
This can also be seen in Fig.~\ref{fig:syre}b that shows how [Fe/O] runs with 
[O/H] as the uprising trend in [Fe/O] at [O/H]\,$\sim$\,0.

%==============================================================================
\subsubsection{Evolution of the thin disk}

The Galactic thin disk has not had such an intense star
formation history as the thick disk. The shallow decline in the [O/Fe] trend
from an oxygen overabundance of [O/Fe]\,$\sim$\,0.2\,--\,0.3 at
[Fe/H]\,$\sim$\,$-0.8$ to an oxygen underabundance of [O/Fe]\,$\approx$\,$-0.2$
at [Fe/H]\,$\sim$\,0.4 
indicates a 
more continuous star formation with no fast initial
enrichment from SN\,II. 
Instead the observed [O/Fe] trend favours a chemical
evolution in the thin disk where both SN\,Ia and SN\,II in a steady rate
contribute to the enrichment of the interstellar medium and thereby producing
a decline in [O/Fe] with [Fe/H] (see Fig.~\ref{fig:syre}a).

How can we explain the trend observed in the thin disk? As noted in the 
preceding section the thin disk stars, on average, are younger than the thick
disk stars. However, the thin disk stars appear to span the same metallicity
range (below solar metallicity) as the thick disk stars do. A possible
scenario for the star formation in the thin disk would be that once star 
formation in the thick disk was stopped (truncated?) there is a pause in the
star formation. During this time there is in-falling fresh gas that
accumulates in the Galactic plane, forming a new thin disk.
Also the remaining gas from the thick disk will settle down onto the
new disk. Once enough material is collected star formation is restarted in
the new thin disk. The gas, though, has been diluted by the metal-poor 
in-falling gas. This means that the first stars to form in the thin disk will 
have lower metallicities than the last stars that formed in the thick disk.
Although this scenario nicely explains the abundance trends in the thin disk
it remains to work out the details and see if they fit other observational
tests such as G dwarf and metallicity distributions.

Prior to this study the models of Galactic chemical evolution did not
match the observed [O/Fe] trend at [Fe/H]\,$>$\,0 since the models predict
a downward trend (see e.g. Chiappini et al.~\cite{chiappini2003} for
the most recent models) while the observed trend leveled out
(see e.g. Nissen et al.~\cite{nissen2002} and discussion in 
Sect.~\ref{sec:strengthening}). Even though the models of
Galactic chemical evolution rarely are evolved beyond [Fe/H]\,$\approx$\,$+0.2$
there is no indication of a leveling out.

%==============================================================================
\subsection{Synthesis of $\alpha$-elements and the choice of 
	reference elements in chemical evolution studies} 
	\label{sect:synthesis}

%------------------------------------------------------------------------------
%------------------------------------------------------------------------------

\begin{figure}
 \resizebox{\hsize}{!}{\includegraphics[bb=18 144 592 475,clip]{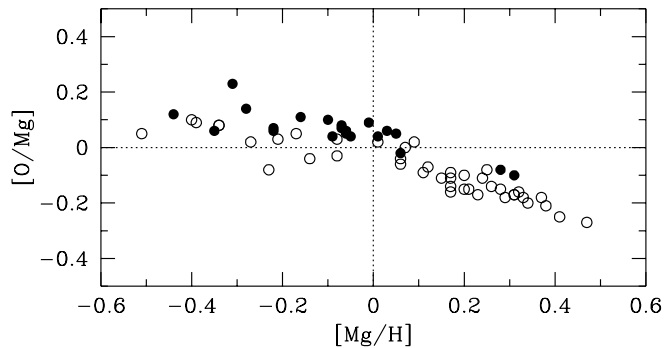}}
\caption{
        [O/Mg] vs. [Mg/H]. Only stars that have oxygen abundances
        from the [\ion{O}{i}]$_{6300}$ line are shown.
        Mg abundances are from Bensby et al.~(\cite{bensby}).
        Thin and thick disk
        stars are marked by open and filled circles, respectively.
         }
\label{fig:alpha}
\end{figure}

The models of Galactic chemical evolution predict that [Mg/Fe], at
[Fe/H]\,$>$\,0, should show a downward trend similar to that seen for
[O/Fe].  Our observations do not show this (Bensby et al.~\cite{bensby}).
Instead for a downward trend [Mg/Fe] levels out at [Fe/H]\,$\approx$\,0.
In Fig.~\ref{fig:alpha} we show [O/Mg] as a function of [Mg/H] which
further emphasizes the different trends for oxygen and Mg.

Prior to the new models by Meynet \& Maeder~(\cite{meynet}) the models by
Maeder~(\cite{maeder}) predicted a strong dependency on the metallicity
for the oxygen yields.  This could then have explained the trend in
Fig.~\ref{fig:alpha}.  But, since the oxygen yields for massive stars
now have been shown to be less sensitive to the metallicity when including
rotation in the stellar models (Meynet \& Maeder~\cite{meynet}) this
explanation is no longer viable.  The stellar evolution models are 
also well-known to underestimate the Mg yields for massive stars (e.g.
Chiappini et al.~\cite{chiappini2003}; Chiappini et
al.~\cite{chiappini1999}; Thomas et al.~\cite{thomas}) and thereby giving
too low [Mg/Fe] values at high [Fe/H] in the models of Galactic chemical
evolution. Currently, this could then be the explanation for the
discrepancy between models of Galactic chemical evolution and observations
for Mg.  However, this conclusion must remain tentative and more work on
stellar yields is necessary for both Mg as well as oxygen in order to
achieve a clear picture. In this our observational data can be used to
further constrain the possible models.

In studies of chemical evolution of single or multiple stellar populations
it is desirable to have a reference element that has only one,
well-understood source. Stellar spectra are rich in easily measurable Fe
lines. This means that for almost all stars Fe abundances are readily
available and hence Fe is by far the most commonly used reference element
in studies of galactic chemical evolution. However, Fe is produced in both
SN\,II and SN\,Ia and therefore the trends are less easy to interpret than
if we had a reference element that has only one source. Such reference
elements could be provided by Mg or oxygen.  However, in light of our own
observations, the shortcomings of stellar yield models, and the fact that
at least Mg might be produced, to some extent, in SN\,Ia we would be very
cautious about promoting one specific reference element. It is probably
best, for now, to consider more than one reference element.

%==============================================================================
\section{Summary} \label{sec:summary}

We have presented oxygen abundances for 72 nearby F and G dwarf stars in the 
solar neighbourhood with $-0.9$\,$<$\,[Fe/H]\,$<$\,$+0.4$. The stellar samples 
were chosen in order to investigate two important issues related to the 
formation and chemical evolution of the stellar disks in our Galaxy:
\begin{itemize}
\item[$\star$] the oxygen trends in the thin and thick disk,
\item[$\star$] the oxygen trend in the thin disk at super-solar metallicities.
\end{itemize}

%------------------------------------------------------------------------------
%------------------------------------------------------------------------------

The stars have been selected so that they either with a high likelihood belong 
to the thick or the thin disk. This selection was purely based on kinematics.
The majority of the stars in the study presented here have previously 
been studied by Bensby et al.~(\cite{bensby}) and 
Feltzing et al.~(\cite{feltzing}). In those two papers we derived stellar 
parameters, detailed statistics of the kinematic selection, and abundances for 
$\alpha$- as well as iron peak-elements.

Our abundance analysis is based primarily on the forbidden oxygen line at 
6300\,{\AA}. This line is blended with a doublet of Ni lines arising from two 
different Ni isotopes. As we have spectra of very high $S/N$ and $R$ and
have modeled the line and its oxygen and Ni components in detail 
we have been able to disclose very well-defined trends for oxygen relative to 
iron and magnesium. We have also analyzed 
the forbidden line at 6363\,{\AA} and the permitted triplet lines around 
7774\,{\AA} and found consistent trends from these lines.

Our main conclusions from this study are:

\begin{itemize}
     \item[{\bf (i)}]
        At super-solar [Fe/H] the [O/Fe] trend for the thin disk continues
        linearly down-wards. This is different from the other $\alpha$-elements
        which show a leveling out of [$\alpha$/Fe] at [Fe/H]\,$=$\,0.

     \item[{\bf (ii)}]
	The thick disk is more overabundant in oxygen than the thin disk
	at sub-solar metallicities. The thick disk also shows
	the signatures of chemical enrichment from SN\,Ia.

     \item[{\bf (iii)}]
	Oxygen and magnesium do not evolve in lockstep at 
	super-solar metallicities.

     \item[{\bf (iv)}]
	By comparing oxygen abundances from the permitted infrared \ion{O}{i}
	triplet to those from the forbidden line at 6300\,{\AA}
	we provide an empirical NLTE-correction relation for the 
	abundances from the triplet lines that.
	This could be used e.g. for F and G dwarf star spectra
	with a S/N that is such that only the triplet
	lines that can be analyzed well, e.g. due to the distances of the 
	stars.

     \item[{\bf (v)}]
        Comparing our abundances from the 6300\,{\AA} line and the triplet
        lines with and without NLTE corrections from
        Gratton et al.~(\cite{gratton2}) we find that their NLTE corrections,
        for the parameter space spanned by our stars, are 
	somewhat overestimated.
\end{itemize}

Conclusion (i) is a new result and indicates that chemical evolution models 
of the Milky Way are now in concordance with the observed [O/Fe] trends
at [Fe/H]\,$>$\,0.
The different trend for oxygen compared to the other $\alpha$-elements
at [Fe/H]\,$>$\,0 (see Bensby et al.~\cite{bensby}) indicates that oxygen is 
{\it only} produced in
SN\,II. The leveling out of the [$\alpha$/Fe] trend for other $\alpha$-elements
could be expected on grounds that they also have small contributions from 
SN\,Ia. However, given the still large uncertainties in SN yield calculations
this conclusion remains tentative.

Conclusion (ii) means that the Galactic thin and the thick disks indeed have
different chemical histories. This is a confirmation and strengthening of 
previous findings by us and others (e.g. Prochaska et al.~\cite{prochaska}). 
We also trace the signature of SN\,Ia (the ``knee") in the [O/Fe] trend in 
the thick disk that we have noticed previously for Ca, Mg, Si, and Ti 
(Bensby et al.~\cite{bensby}; Feltzing et al.~\cite{feltzing}).

%==============================================================================
\acknowledgements

We would like to thank 
Bengt Gustafsson, Kjell Eriksson, Bengt Edvardsson, and Martin Asplund
for letting us use the Uppsala MARCS code. We further thank BG, KE and BE       
for the use of the {\sc Eqwidth} abundance program and the 
{\sc Spectrum} line synthesis program. 
We would also like to thank Poul Erik Nissen, Bengt Gustafsson, and the 
anonymous referee for valuable comments on the submitted manuscript.
TB also thanks 
Kungliga Fysiografiska S\"allskapet i Lund for financial support to the first 
trip to La Silla in September 2000 and SF is grateful for computer grants from 
the same society. This research has made use of the SIMBAD database, operated 
at CDS, Strasbourg, France.

%==============================================================================


\begin{thebibliography}{}

\bibitem[1989]{abia}
 Abia, C., \& Rebolo, R., 1989, ApJ, 347, 186

\bibitem[2001]{allendeprieto}
 Allende Prieto, C., Lambert, D.L., \& Asplund, M., 2001, ApJ, 556, L63

\bibitem[1994]{andersson}
 Andersson, H., \& Edvardsson, B., 1994, A\&A, 290, 590

\bibitem[1995]{anstee}
 Anstee, S.D., \& O'Mara, B.J., 1995, MNRAS, 276, 859

\bibitem[2003]{asplund}
 Asplund, M., 2003, ASP Conf. Ser., vol. 304, CNO in the universe, 
 eds. C. Charbonnel, D. Schaerer, \& G. Meynet,
 10-14 September 2002 @ Saint-Luc, Valais, Switzerland, in press (astro-ph/0302409)

\bibitem[1997]{asplund2}
 Asplund, M., Gustafsson, B., Kiselman, D., \& Eriksson, K., 1997, A\&A, 318, 
 521

\bibitem[2001]{balachandran}
 Balachandran, S.C., Carr, J.S., \& Carney, B.W., 2001, NewAR, 45, 529

\bibitem[1989]{barbuy}
 Barbuy, B., \& Erdelyi-Mendes, M., 1989, A\&A, 214, 239

\bibitem[2001]{barbuy2}
 Barbuy, B., \& Nissen, P.E., Peterson, R.C., \& Spite, F., 2001, 
 New Astronomy Reviews, 45, 509

\bibitem[1997]{barklem}
 Barklem, P.S., \& O'Mara, B.J., 1997, MNRAS, 290, 102

\bibitem[1998]{barklem3}
 Barklem, P.S., \& O'Mara, B.J., 1998, MNRAS, 300, 863

\bibitem[1998]{barklem2}
 Barklem, P.S., O'Mara, B.J., \& Ross, J.E., 1998, MNRAS, 296, 1057

\bibitem[2000]{barklem4}
 Barklem, P.S., Piskunov, N., \& O'Mara, B.J., 2000, A\&AS, 142, 467

\bibitem[2003]{bensby}
 Bensby, T., Feltzing, S., \& Lundstr\"om, I., 2003, A\&A, 410, 527

\bibitem[1991]{bessel}
 Bessel, M.S., Sutherland, R.S., \& Ruan, K., 1991, ApJ, 383, L71

\bibitem[1992]{biemont}
 Bi\'emont, E., \& Zeippen, C.J., 1992, 265, 850

\bibitem[1999]{boesgaard}
 Boesgaard, A.M., King, J.R., Deliyannis, C.P., \& Vogt, S.S., 1999, AJ, 117, 
 492

\bibitem[2000]{carretta}
 Carretta, E., Gratton, R.G., \& Sneden, C., 2000, A\&A, 356, 238

\bibitem[1997]{castro}
 Castro, S., Rich, R.M., Grenon, M., Barbuy, B., \& McCarthy, J.K., 1997, 
 AJ, 114, 376

\bibitem[2000]{chen}
 Chen, Y.Q., Nissen, P.E., Zhao, G., Zhang, H.W., \& Benoni, T., 2000, A\&AS, 
 141, 491

\bibitem[2003]{chen2003}
 Chen, Y.Q., Zhao, G., Nissen, P.E., Bai, G.S., \& Qiu, H.M., 2003, ApJ, 
 591, 925

\bibitem[1999]{chiappini1999}
 Chiappini, C., Matteucci, F., Beers, T.C., \& Nomoto, K., 1999, ApJ, 515, 226

\bibitem[2003]{chiappini2003}
 Chiappini, C., Romano, D., \& Matteucci, F., 2003, MNRAS, 339, 63

\bibitem[1981]{clegg}
 Clegg, R.E.S., Lambert, D.L., \& Tomkin, J., 1981, ApJ, 250, 262

\bibitem[2003]{lide}
 CRC Handbook of Chemistry and Physics 83rd edition, 2003, Lide, D.R. (ed.), 
 CRC Press

\bibitem[2002]{dalcanton}
 Dalcanton, J.J., \& Bernstein, R.A., 2002, AJ, 124, 1328

\bibitem[1998]{laverny}
 de Laverny, P., \& Gustafsson, B., 1998, A\&A, 332, 661

\bibitem[1993]{edvardsson}
 Edvardsson, B., Andersen, J., Gustafsson, B., Lambert, D.L., Nissen, P.E.,
 \& Tomkin, J., 1993, A\&A, 275, 101

\bibitem[1979]{eriksson}
 Eriksson, K., \& Toft, S.C., 1979, A\&A, 71, 178

\bibitem[1998]{feltzing1998}
 Feltzing, S., \& Gustafsson, B., 1998, A\&AS, 129, 237

\bibitem[2003]{feltzing}
 Feltzing, S., Bensby, T., \& Lundstr\"om, I., 2003, A\&A, 397, L1

\bibitem[1986]{gratton1986}
 Gratton , R.G., \& Ortolani, S., 1986, A\&A, 169, 201

\bibitem[1999]{gratton2}
 Gratton, R.G., Carretta, E., Eriksson, K., \& Gustafsson, B., 1999, A\&A, 
 350, 955

\bibitem[2000]{gratton}
 Gratton, R.G., Carretta, E., Matteucci, F., \& Sneden, C., 2000, A\&A, 
 358, 671

\bibitem[2003]{gratton2003}
 Gratton, R.G., Carretta, E., Desidera, S., Lucatello, S., Mazzei, P., 
 \& Barbieri, M., 2003, A\&A, 406, 131

\bibitem[1998]{grevesse}
 Grevesse, N., \& Sauval, A.J., 1998, Space Sci. Rev., 85, 161

\bibitem[1975]{gustafsson}
 Gustafsson, B., Bell, R.A., Eriksson, K., \& Nordlund, {\AA}., 1975, A\&A, 
 42, 407

\bibitem[1999]{gustafsson2}
 Gustafsson, B., Karlsson, T., Olsson, E., Edvardsson, B., \& Ryde, N., 
 1999, A\&A, 342, 426

\bibitem[1991]{hibbert}
 Hibbert, A., Bi\'emont, E., Godefroid, M., \& Vaeck, N., 1991, 
 J. Phys. B: At. Mol. Opt. Phys., 24, 3943

\bibitem[1998]{israelian2}
 Israelian, G., Garc\'ia L\'opez R.J., \& Rebolo, R., 1998, ApJ, 507, 805

\bibitem[2001]{israelian}
 Israelian, G., Rebolo, R., Garc\'ia L\'opez R.J., et al., 2001, ApJ, 551, 833

\bibitem[2003]{johansson}
 Johansson, S., Litz\'en, U., Lundberg, H., \& Zhang, Z., 2003, ApJ, 584, L107

\bibitem[1991]{kiselman}
 Kiselman, D., 1991, A\&A, 245, L9

\bibitem[1993]{kiselman2}
 Kiselman, D., 1993, A\&A, 275, 269

\bibitem[1982]{kjaergaard}
 Kj\ae rgaard, P., Gustafsson, B., Walker, G.A.H., \& Hultqvist, L., 1982, 
 A\&A, 115, 145

\bibitem[2002]{kroupa}
 Kroupa, P., 2002, MNRAS, 330, 707

\bibitem[1999]{kupka} 
 Kupka, F., Piskunov, N., Ryabchikova, T.A., Stempels, H.C., \& Weiss, W.W., 
 1999, A\&AS, 138, 119

\bibitem[2001]{kurster}
 K\"urster, M., 2001, The CES User Manual, Version 1.0, European Southern 
 Observatory (ESO)

\bibitem[1978]{lambert}
 Lambert, D.L., 1978, MNRAS, 182, 249

\bibitem[1974]{lambert1974}
 Lambert, D.L., Sneden, C., \& Ries, L.M., 1974, ApJ, 188, 97

\bibitem[1975]{mackle}
 M\"ackle, R., Holweger, H., Griffin, R., \& Griffin, R., 1975, A\&A, 38, 239

\bibitem[1986]{matteucci}
 Matteucci, F., \& Greggio, L., 1986, A\&A, 154, 279

\bibitem[1992]{maeder}
 Maeder, A., 1992, A\&A, 264, 105

\bibitem[2002]{melendez}
 Mel\'endez, J., \& Barbuy, B., 2002, ApJ, 575, 474

\bibitem[2001]{melendez2}
 Mel\'endez, J., Barbuy, B., \& Spite, F., 2001, ApJ, 556, 858

\bibitem[2003]{meynet}
 Meynet, G., \& Maeder, A., 2003, A\&A, 390, 561

\bibitem[2000]{mishenina}
 Mishenina, T.V., Korotin, S.A., Klochkova, V.G., \& Panchuk, V.E., 2000, 
 A\&A, 353, 978

\bibitem[1965]{mitchell}
 Mitchell, W.E. Jr., \& Mohler, O.C., 1965, ApJ, 141, 1126

\bibitem[1966]{moore}
 Moore, C.E., Minnaert, M.G.J., \& Houtgast, J., 1966, 
 The Solar Spectrum 2935 {\AA} to 8770 {\AA},
 National Bureau of Standards  Monograph 61

\bibitem[1994]{nave}
 Nave, G., Johansson, S., Learner, R.C.M., Thorne, A.P., \& Brault, J.W., 1994,
 ApJS, 94, 221

\bibitem[1951]{naqvi}
 Naqvi, A.M., 1951, Thesis Harvard

\bibitem[1997]{schuster}
 Nissen, P.E., \& Schuster, W.J., 1997, A\&A, 326, 751

\bibitem[1992]{nissen1992}
 Nissen, P.E., \& Edvardsson, B., 1992, A\&A, 261, 255

\bibitem[2002]{nissen2002}
 Nissen, P.E., Primas, F., Asplund, M., \& Lambert, D.L., 2002, A\&A, 390, 235

\bibitem[1960]{omholt}
 Omholt, A., 1960, Planetary and Space Science, 2, 246

\bibitem[1995]{piskunov}
 Piskunov, N.E., Kupka, F., Ryabchikova, T.A., Weiss, W.W., \& Jeffery, C.S., 
 1995, A\&AS, 112, 525

\bibitem[2002]{pompeia}
 Pomp\'eia, L., Barbuy, B., \& Grenon, M., 2002, ApJ, 566, 845

\bibitem[2000]{prochaska}
 Prochaska, J.X., Naumov, S.O., Carney, B.W., McWilliam, A., \& Wolfe, A.M., 
 2000, ApJ, 120, 2513

\bibitem[1993]{quinn}
 Quinn, P.J., Hernquist, L., \& Fullagar, D.P., 1993, ApJ, 403, 74

\bibitem[2003]{reddy}
 Reddy, B.E., Tomkin, J., Lambert, D.L., \& Allende Prieto, C., 
 2003, MNRAS, 340, 304

\bibitem[1997]{reshetnikov}
 Reshetnikov, V., \& Combes, F., 1997, A\&A, 324, 80

\bibitem[1999]{ryabchikova}
 Ryabchikova, T.A., Piskunov, N.E., Stempels, H.C., Kupka, F., \& Weiss W.W., 
 1999, Proc. of the 6th International Colloquium on Atomic Spectra and 
 Oscillator Strengths, Victoria BC, Canada, 1998, Physica Scripta, T83, 162 
 (VALD-2) 

\bibitem[2000]{schwarzkopf}
 Schwarzkopf, U., \& Dettmar, R.-J., 2000, A\&A, 361, 451

\bibitem[2002]{shi}
 Shi, J.R., Zhao, G., \& Chen, Y.Q., 2002, A\&A, 381, 982

\bibitem[1979]{sneden}
 Sneden, C., Lambert, D.L., \& Whitaker, R.W., 1979, ApJ, 234, 964

\bibitem[1960]{stoffregen}
 Stoffregen, W., \& Derblom, H., 1960, Nature, 185, 28

\bibitem[2001]{tautvaisiene}
 Tautvai\v sien\.e, G., Edvardsson, B., Tuominen, I., \& Ilyin, I., 
 2001, A\&A, 380, 579

\bibitem[1998]{thomas}
 Thomas, D., Greggio, L., \& Bender, R., 1998, MNRAS, 296, 119

\bibitem[1979]{tinsley}
 Tinsley, B.M., 1979, ApJ, 229, 1046

\bibitem[1989]{wheeler}
 Wheeler, J.C., Sneden, C., \& Truran Jr., J.W., 1989, ARAA, 27, 279

\bibitem[1966]{wiese}
 Wiese, W.L., Smith, M.W., \& Glennon, B.M., 
 1966, Atomic Transition Probabilities, Vol. 1 
 Hydrogen Through Neon, National Bureau of Standards 4 (NSRDS-NBS 4)

\bibitem[1952]{yamanouchi}
 Yamanouchi, T., \& Horie, H., 1952, J. Phys. Soc. Japan, 7, 52


\end{thebibliography}
\end{document}